\documentclass[10pt,journal,compsoc]{IEEEtran}
\usepackage{amsmath, amsfonts, bm}
\usepackage[switch]{lineno}
\usepackage{CJKutf8}
\usepackage{multirow}
\usepackage{subcaption}
\usepackage{graphicx}
\usepackage{makecell}
\usepackage{xcolor}
\usepackage{ragged2e}
\usepackage{fancyhdr}
\usepackage{url}
\usepackage[colorlinks=true,urlcolor=blue]{hyperref}
\newcommand{\beginsupplement}{
        \setcounter{table}{0}
        \renewcommand{\thetable}{\arabic{table}}
        \setcounter{figure}{0}
        \renewcommand{\thefigure}{\arabic{figure}}
     }
     
\usepackage{float}
\ifCLASSOPTIONcompsoc
  \usepackage[nocompress]{cite}
\else
  \usepackage{cite}
\fi

\ifCLASSINFOpdf
\else
\fi
    
\begin{document}

\newpage
\thispagestyle{empty}
\onecolumn

\noindent{\fontsize{11}{14}\selectfont This is the accepted manuscript of the journal article, published in IEEE Transactions on Affective Computing. The citation is `Z. Li \textit{et al}., "Towards human-compatible autonomous car: A study of non-verbal Turing test in automated driving with affective transition modelling," in \textit{IEEE Transactions on Affective Computing}, doi: \href{https://doi.org/10.1109/TAFFC.2023.3279311}{10.1109/TAFFC.2023.3279311}'.}

\vspace{12pt}

\noindent{\fontsize{11}{14}\selectfont © 2023 IEEE. Personal use of this material is permitted. Permission from IEEE must be obtained for all other uses, in any current or future media, including reprinting/republishing this material for advertising or promotional purposes, creating new collective works, for resale or redistribution to servers or lists, or reuse of any copyrighted component of this work in other works.}

\newpage

\twocolumn

\clearpage
\pagenumbering{arabic}
\setcounter{page}{1}

\title{Towards human-compatible autonomous car: A study of non-verbal Turing test in automated driving with affective transition modelling}

\author{Zhaoning~Li,
        Qiaoli~Jiang,
        Zhengming Wu,
        Anqi~Liu,
        Haiyan~Wu,~\IEEEmembership{Member,~IEEE,}
        Miner Huang,
        Kai Huang,
        and~Yixuan~Ku
\IEEEcompsocitemizethanks{
\IEEEcompsocthanksitem Z. Li, Q. Jiang, M. Huang and Y. Ku are with the Guangdong Provincial Key Laboratory of Brain Function and Disease, Centre for Brain and Mental Well-being, Department of Psychology, Sun Yat-sen University, Guangzhou, China.\protect\\
E-mail: lizhn7@mail2.sysu.edu.cn, jiangqli@mail2.sysu.edu.cn,\\ 
edshme@mail.sysu.edu.cn, kuyixuan@mail.sysu.edu.cn\\
Twitter: @lizhn7; Mastodon: @lizhn7@sciences.social
\IEEEcompsocthanksitem Z. Li and H. Wu are with the Centre for Cognitive and Brain Sciences and Department of Psychology, University of Macau, Taipa, Macau, China.\protect\\
E-mail: yc17319@umac.mo, haiyanwu@um.edu.mo
\IEEEcompsocthanksitem Z. Wu is with the Guangzhou Intelligent Connected Vehicle Pilot Zone Operations Centre, Guangzhou, China.\protect\\
E-mail: 13590335095@163.com
\IEEEcompsocthanksitem A. Liu is with the Department of Computer Science, Whiting School of Engineering, Johns Hopkins University, Baltimore, United States.\protect\\
E-mail: aliu@cs.jhu.edu
\IEEEcompsocthanksitem K. Huang is with the School of Computer Science and Engineering, Sun Yat-Sen University, Guangzhou, China.\protect\\
E-mail: huangk36@mail.sysu.edu.cn
\IEEEcompsocthanksitem Y. Ku is the corresponding author.
}}

\IEEEtitleabstractindextext{%

\begin{abstract}

Autonomous cars are indispensable when humans go further down the hands-free route. Although existing literature highlights that the acceptance of the autonomous car will increase if it drives in a human-like manner, sparse research offers the naturalistic experience from a passenger's seat perspective to examine the humanness of current autonomous cars. The present study tested whether the AI driver could create a human-like ride experience for passengers based on 69 participants' feedback in a real-road scenario. We designed a ride experience-based version of the non-verbal Turing test for automated driving. Participants rode in autonomous cars (driven by either human or AI drivers) as a passenger and judged whether the driver was human or AI. The AI driver failed to pass our test because passengers detected the AI driver above chance. In contrast, when the human driver drove the car, the passengers' judgement was around chance. We further investigated how human passengers ascribe humanness in our test. Based on Lewin's field theory, we advanced a computational model combining signal detection theory with pre-trained language models to predict passengers' humanness rating behaviour. We employed affective transition between pre-study baseline emotions and corresponding post-stage emotions as the signal strength of our model. Results showed that the passengers' ascription of humanness would increase with the greater affective transition. Our study suggested an important role of affective transition in passengers' ascription of humanness, which might become a future direction for autonomous driving.
\end{abstract}

} 

\maketitle

\begin{quote}
`Well, I'm human in part.'

\ ... `Which part, Andrew?'

\ ... `My mind. My heart. I may be artificial, alien, inhuman so far as your strict genetic definition goes. But I’m human in every way that counts. And I can be recognised as such legally.'

\qquad\qquad\qquad \ \ \ Isaac Asimov and Robert Silverberg

\qquad\qquad\qquad\qquad\qquad\quad \ \ \ \ \ The Positronic Man \cite{asimov1992positronic}\\\\

\end{quote}

\ifCLASSOPTIONcompsoc
\IEEEraisesectionheading{\section{Introduction}}
\else
\section{Introduction}
\fi

\IEEEPARstart{1}{,350,000}, a heart-breaking statistics number, is from the global status report on road safety 2018 \cite{world2018global}, which means the number of deaths from road traffic crashes worldwide. In other words, every 23 seconds, there will be someone in this world to be told that a loved one has died in a road crash. It is worth noting that 94\% of road accidents are due to human error \cite{nhtsa}. Based on these critical and tragic facts, it is promising that autonomous cars \footnote{According to the literature \cite{payre2020although}, which promotes using the term `autonomous cars' to facilitate public acceptance of automated driving from a range of terms, we adopted the term `autonomous cars' in this paper for this endeavour.}(ACs) have the potential to reduce human error substantially \cite{taeihagh2019governing,rojas2020autonomous}. Notably, artificial intelligence (AI) algorithms in ACs can make faster driving decisions than human drivers to prevent crashes \cite{lim2019algorithmic,Mohammed_2019}. Globally, ACs are poised to save 10 million lives simply by removing human error elements per decade \cite{fleetwood2017public}.

Despite the lifesaving benefits of ACs being paramount, and researchers from academia and industry have made significant progress \cite{kato2018autoware,chishiro2019towards,ap} since the first landmark AC appeared nearly 40 years ago \cite{wallace1985first,kanade1986autonomous,turk1988vits}, there has yet to be a large-scale deployment of ACs \cite{Carpenter_2020,Emuna2020}. In other words, ACs still face enormous challenges in replacing humans (e.g., Tesla Autopilot deaths \cite{Tesla}). In addition to safety and trust issues \cite{merriman2021challenges,merriman2021can,manchon2021calibration}, another main obstacle is that these cars are not humanoid, which means they are not driving in a human-like manner. More importantly, existing literature highlights that the acceptance of the AC will increase if the AC drives in a human-like manner \cite{al2001framework,al2003toward,gu2017human,sun2020exploring} (the rationale is that `humans will find it easier to interact and feel at ease with ACs in such cases \cite{hecker2019learning}').

In this regard, a variety of algorithmic researchers have proposed sophisticated algorithms concerning human-like driving trajectories based on estimated human behaviour or perceived risk \cite{kraus2010optimisation,guo2018toward,shin2018human,xu2020learning,wang2020human,kolekar2020human}, human-like decision-making at intersections \cite{de2017decision}, human-like car following \cite{fu2019human}, human-like braking behaviour \cite{lehsing2019don}, human-like crawling forward at pedestrian crossings \cite{waymo}, human-like peeking when approaching road junctions \cite{oliveira2019driving}, human-like cost function \cite{liu2019modeling} and human-like driving policies in collision avoidance \cite{Emuna2020,Li_2021} and merging \cite{Schwarting_2019} to increase the human likeness of ACs, i.e., teaching ACs about human-like driving from the algorithmic perspective. 

On the other hand, many researchers in the area of human factors mainly use simulators in the laboratory arena or online surveys to examine how drivers \cite{griesche2016should,hartwich2018driving,sun2020exploring,Stanton2020}, pedestrians \cite{fuest2018using,bazilinskyy2021driving} or passengers \cite{rossner2019you,oliveira2019driving,dettmann2021comfort} respond to ACs, which have been programmed and designed deliberately to perform in a human-like manner, i.e., validating algorithms of ACs from the perspective of human factors studies. However, sparse research offers a true-to-life ride experience for passengers to examine the humanness of the AC. Given that it is the key to improving the acceptance of the AC, we presented the following research question:

\textbf{How to offer the naturalistic experience from a passenger's seat perspective to measure the humanness of current ACs?}

To tackle the research question and overcome the limitations of driving simulators and laboratory settings, we developed a ride experience-based version of the `Turing test \cite{Turing1950}' on SAE Level 4 \cite{SAEInternational2018} ACs (in which human intervention is unnecessary in limited spatial areas or under special circumstances) in the real world (Section~\ref{sec2.1}). In 1950, Alan Turing proposed the Turing test \cite{Turing1950} to evaluate the ascription of intelligence, i.e., whether humans would ascribe human-like intelligent behaviour to machines. In the Turing test, a human interrogator (C) engages in verbal interaction with a computer program (A) and another human (B) (i.e., asks questions to A and B through written notes) and tries to determine with whom C is interacting. If the human interrogator cannot determine which answers are given by a human and which by a computer program, the latter is said to pass the test. The rationale of the Turing test, i.e., `human judges impartially compare and evaluate outputs from different systems while ignoring the source of the outputs \cite{moor2001status}', has been used in the literature \cite{pfeiffer2011non,ciardo2022human} in different ways to investigate the ascription of humanness. Based on the similar spirit of the above work, Cascetta et al. \cite{CASCETTA2022103499} proposed a Turing test approach to investigate the humanness of SAE Level 2 ACs under different traffic conditions and driving actions on a real road circuit. Emuna et al. \cite{Emuna2020} and Zhang et al. \cite{https://doi.org/10.1002/aisy.202100211} also performed variations of the Turing test to assess the humanness of their proposed self-driving algorithms using either pre-recorded video or driving simulators.

\begin{figure*}[!t]
\center
\includegraphics[width=\textwidth]{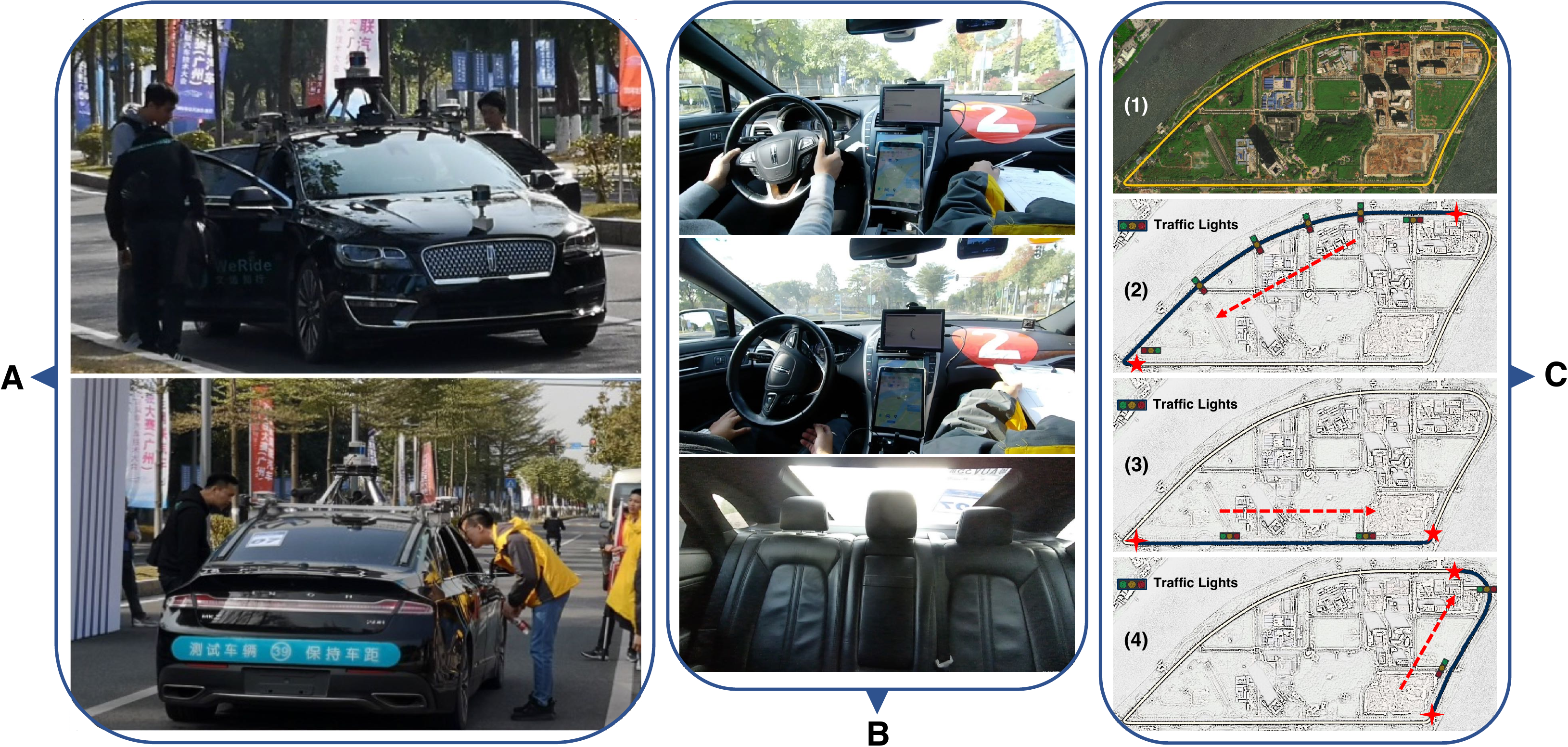}
\caption{The non-verbal variation of the Turing test for automated driving in detail. \textbf{A.} SAE Level 4 AC used in the test. \textbf{B.} From top to bottom: manual driving mode (the human driver was actively steering); autonomous mode, in which the human driver would be free to release the steering wheel, meaning that the AI driver (i.e., WeRide ONE algorithm \cite{Weride}) would take control of the car; the participants would ride in the rear seat taking the role of a passenger, and a thick black drape hid the driver cabin from the passenger's viewpoint. \textbf{C.} Sub-figure (1) is the satellite image of the test stages (yellow colour). The dark blue of the sub-figure (2-4) represents the first, second and third stages, respectively. Each participant would experience three stages in turn (randomly assigned to the manual driving mode or autonomous mode). The red arrow indicates the direction of travel. The 4-point star and 5-point star represent the start and end location of the AC, respectively. Furthermore, the traffic lights in all three stages have been marked.}
\label{fig1}
\end{figure*}

Within the framework of the Turing test, the humanness investigated in this paper is intended as the ability of ACs to create a ride experience for passengers that is not distinguishable from that created by human drivers. To that end, we conducted a ride experience-based version of the Turing test (Section~\ref{sec2.1.2}), which in the following will be referred to as `a non-verbal variation of the Turing test'. In particular, 69 participants (Section~\ref{sec2.1.1}) assumed the role of a passenger in the rear seat of SAE Level 4 cars, unable to see the driver cabin, as depicted in Fig.~\ref{fig1}. They had to judge whether a real human or an AI algorithm was behind the wheel based on their ride experience on the road stage just passed. Specifically, after each road stage, passengers rated a variation of a Turing test question, `Do you think the driver was a real human or an AI algorithm?', with 1-3, 1 for \textit{`AI driver'}; 2 for \textit{`Not sure'}; 3 for \textit{`Human driver'}. Our main goal was to test whether the AI driver (i.e., WeRide ONE, a universal self-driving algorithm for comprehensive open urban roads \cite{Weride}) could create a human-like ride experience for passengers, such that passengers would have either chance-level or even higher humanness ratings under the AI driver condition. The results showed that when the AI driver controlled the AC, passengers' humanness ratings were significantly below the chance level, indicating that passengers could detect and discriminate between the human and AI drivers. Thus, the AI driver did not pass our non-verbal variation of the Turing test (Section~\ref{sec3.1}).

The AI driver's failure inspired us to explore further why the AI algorithm could trick human passengers in some trials and not in most others. Digging into this rabbit hole may bring us some informative information for future ACs. Accordingly, we presented the following thought-provoking research question:

\textbf{How do human passengers ascribe humanness in the non-verbal variation of the Turing test?}

In our non-verbal variation of the Turing test, passengers' ascription of humanness would likely be influenced by their affective states, cognitive inference and external stimuli (i.e., human and AI drivers). In particular, we could further formulate the problem mentioned above in the language of field theory \cite{lewin1936}, the most central and influential work of Kurt Lewin \cite{burnes2013k} (who is undoubtedly the father of modern social psychology). Field theory states that a person's psychological field (i.e., the total psychological environment that the person experiences subjectively) determines their behaviour ($B$) \cite{lewin1951field}, which can be expressed by Lewin's equation \cite{2008}:

\begin{equation}
\label{equ1}
B = f(P,E)
\end{equation}

\begin{figure*}[!t]
\center
\includegraphics[width=\textwidth]{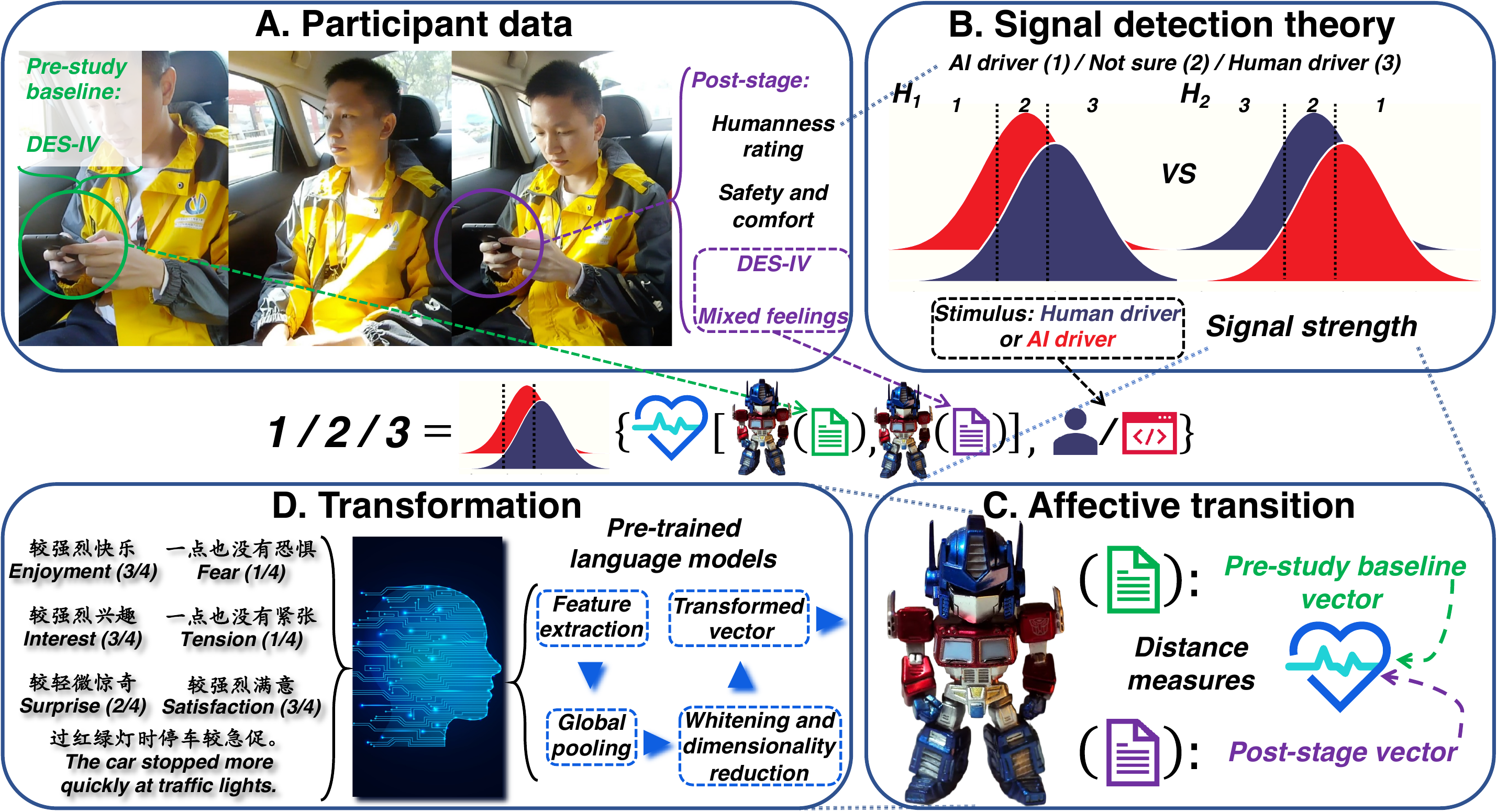}
\caption{Schematic illustration of the computational modelling. The modelling process was underpinned by Lewin's field theory \cite{lewin1936}, which is expressed by a cartoon version of the formula: $B = f(P,E)$ at the centre of the figure. \textbf{A.} From left to right: A participant was filling out pre-study self-reported scores of the modified DES-IV on his smartphone; the stage began; after the stage, the participant was completing the online questionnaire, including his humanness rating (i.e., the answer to a variation of a Turing test question, `Do you think the driver was a real human or an AI algorithm?', 1 for \textit{`AI driver'}; 2 for \textit{`Not sure'}; 3 for \textit{`Human driver'}), post-stage modified DES-IV scores, the scores of safety and comfort, and written mixed feelings (optional). \textbf{B.} The high-level illustration of our model, with the framework of SDT as the backbone. The signal strength (computed as affective transition, AT) and stimulus (human driver or AI driver) are the model's input, while the output is the participant's humanness rating behaviour. Notice that the two competing hypotheses ($\textit{H}_{1}$ and $\textit{H}_{2}$) about the possible relatedness between the participant's humanness rating and the magnitude of signal strength are all depicted. \textbf{C.} The further computation of AT, i.e., the computation of the distance between the pre-study baseline and post-stage vectors, in which vectors would be transformed by Optimus Prime (a fictional character created by the Transformers franchise), i.e., the transformation module. \textbf{D.} The internal transformation procedure when giving Optimus Prime a participant's post-stage rating scores and mixed feelings.}
\label{fig2}
\end{figure*}

where $P$ and $E$ represent the person and their environment, respectively. In line with the idea of Gestalt psychology that `the whole is more than the sum of its parts \cite{sternberg2013cognitive}', the parts, i.e., $P$ and $E$, together combine to form something larger, the psychological field \cite{marrow1977practical}. In our case, $B$ is humanness rating behaviour, and $P$ and $E$ denote passenger and the driving environment, respectively. Given $B$, we aimed to figure out the computation of the right-hand side of Lewin's equation. To phrase the matter another way, we intended to investigate why a given trial (i.e., a particular passenger $P$ in a particular driving environment $E$) has event $B$ (e.g., high humanness rating) and no other as a result of the non-verbal variation of the Turing test. To that end, we proposed a computational model which combines signal detection theory (SDT) \cite{marcum1947statistical,tanner1954decision,green1966signal} with pre-trained language models (PLMs) \cite{P18-2023,clark2020electra,raffel2020exploring} (Section~\ref{sec2.2}), as depicted in Fig.~\ref{fig2}. In this SDT-based model (Section~\ref{sec2.2.2}), we used affective transition (AT, Section~\ref{sec2.2.3}) between pre-study baseline emotions and corresponding post-stage emotions (collected using the modified Differential Emotions Scale and written description, Section~\ref{sec2.2.1}), transformed by PLM (Section~\ref{sec2.2.4}), as the signal strength.

The results showed that our proposed computational model could adequately predict passengers' humanness rating behaviour in the non-verbal variation of the Turing test (Section~\ref{sec3.2}). Further analysis (Section~\ref{sec4}) suggested that affective transition, serving as a hypothetical essential part (i.e., $P$) of passengers' subjective ride experience in our model, may play a crucial role in their ascription of humanness. Specifically, we found that the passengers' ascription of humanness would increase with the greater AT (Section~\ref{sec4.1}). Moreover, based on the analysis of AT, we also gave concrete suggestions for the AI driver to offer a human-like ride experience for the passenger (Section~\ref{sec4.2}-3). Taking the results of behavioural experiments and computational modelling together, we conjecture that the lack of a certain level of mentalising ability in the current self-driving algorithm may underlie its failure to pass our non-verbal variation of the Turing test. In this regard, our study calls for a spotlight on the importance of ensuring ACs (or artificial social intelligence, more broadly speaking) have at least some mentalising ability (Section~\ref{sec5}).

\section{Methods \label{sec2}}

\subsection{The non-verbal variation of the Turing test \label{sec2.1}}

In this subsection, we will first brief on the information of participants and how we recruit them. Next, we give the details of the conducted non-verbal variation of the Turing test, as illustrated in Fig.~\ref{fig1}.

\subsubsection{Participants \label{sec2.1.1}}

We recruited 23 employees of WeRide (a Chinese high-tech company aiming to develop the most advanced autonomous driving technology) and 46 tourists and passers-by via on-site registration in the Guangzhou International Biological Island. The entire sample included 45 males and 24 females, aged 34.48 (\textit{SD} = 10.44, range = $[21,60]$) years on average. After welcoming, all participants received information about the aims of the experiment and provided informed consent. Participants each received a plush toy for participating in our study. The local ethics committee approved our research protocol (2020-0515-0140).

\subsubsection{Procedure \label{sec2.1.2}}

In the double-blind non-verbal variation of the Turing test, participants went by the SAE Level 4 AC (see Fig.~\ref{fig1}A), driven by either the human driver (manual driving mode) or the AI driver (autonomous mode). Due to legal restrictions, one engineer would place in the passenger's seat (the front one) to monitor the AC. Therefore, the participants would ride in the rear seat, taking the role of a passenger, with a thick black drape hiding the driver cabin from the passenger's viewpoint (see Fig.~\ref{fig1}B). If the passenger cannot distinguish between the manual driving mode and autonomous mode, the non-verbal variation of the Turing test is passed. In turn, each participant would experience three stages assigned randomly to the manual driving mode or autonomous mode.

There were three stages in a 3.4-km-long real road circuit \footnote{Although the road circuit is a real road, it is important to note that the environment may only partially capture the complexity and variability of real-life road situations. The car used in the experiment had a safety prompt, which is not typically present in daily real-world scenarios.} over the Guangzhou International Biological Island in Panyu District, Guangzhou, Guangdong Province, China. Starting from the Xingdao Ring Road North, the first stage (around 1.6 km) included six traffic lights and a left-hand turn towards the second stage on Luoxuan Avenue. After a straight ride of around 1.2 km with two traffic lights, a left-hand turn to the third stage on the Xingdao Ring Road South was performed. Finally, the third stage is around 0.6 km, including a big left-hand curve and two traffic lights. The predetermined course ended at the beginning of the first stage (see Fig.~\ref{fig1}C). Self-reported emotions were assessed before the whole study as pre-study baseline emotions. In contrast, humanness ratings, safety, comfort, post-stage emotions and mixed feelings were measured during the lag time (participants have 1-2 minutes to rest before the next stage) after each stage, and we will further introduce these data in Section~\ref{sec2.2.1}.

\subsection{How do human passengers ascribe humanness? \label{sec2.2}}

To understand passengers' ascription of humanness in the non-verbal variation of the Turing test, we advanced a computational model which specifies the detailed steps for generating passengers' humanness rating behaviour, as shown in Fig.~\ref{fig2}. At the centre of Fig.~\ref{fig2}, we portrayed Lewin's equation \cite{lewin1936}, $B = f(P, E)$, for the highest-level illustration of our computational modelling method. In the following four parts, we will first introduce the details of the participant data collected in the non-verbal variation of the Turing test (see Fig.~\ref{fig2}A) and subsequently describe our model in detail from a top-down perspective (see Fig.~\ref{fig2}B-D).

\subsubsection{Participant data: Self-reported scores, humanness ratings, and mixed feelings \label{sec2.2.1}}

We collected self-reported scores (including pre-study baseline emotions, post-stage emotions, safety and comfort), humanness ratings and mixed feelings from participants in the non-verbal variation of the Turing test (Fig.~\ref{fig2}A). Specifically, pre-study baseline emotions and post-stage emotions were collected using the modified DES-IV \cite{izard1993stability,huang2001} (on Likert scales from 1-4) since it has been suggested that passengers' emotion plays a fundamental role in the social acceptance of ACs \cite{Carpenter_2020,sini2020automatic,marceddu2021novel}. The left side of Fig.~\ref{fig2}D shows an example in which a participant rated six emotions as follows: `\begin{CJK*}{UTF8}{gkai}较强烈快乐\end{CJK*}' (Enjoyment 3/4), `\begin{CJK*}{UTF8}{gkai}较强烈兴趣\end{CJK*}' (Interest 3/4), `\begin{CJK*}{UTF8}{gkai}较轻微惊奇\end{CJK*}' (Surprise 2/4), `\begin{CJK*}{UTF8}{gkai}一点也没有恐惧\end{CJK*}' (Fear 1/4), `\begin{CJK*}{UTF8}{gkai}一点也没有紧张\end{CJK*}' (Tension 1/4), `\begin{CJK*}{UTF8}{gkai}较强烈满意\end{CJK*}' (Satisfaction 3/4).

Besides, user acceptance also resides in the increase of their trust towards the AC \cite{Dimitrakopoulos_2020}. Therefore, in the light of passengers' safety and comfort could establish trust towards the AC \cite{Fraedrich_2016,nastjuk2020drives,Othman_2021}, self-reported scores of safety and comfort were rated on an integer scale from 1 to 4, 1 meaning `Not safe (comfortable) at all' and 4 meaning `Very safe (comfortable)'. Besides, the humanness rating behaviour ($B$ in Lewin's equation), i.e., the answer to a variation of a Turing test question, `Do you think the driver was a real human or an AI algorithm?', was rated from 1-3, 1 for \textit{`AI driver'}; 2 for \textit{`Not sure'}; 3 for \textit{`Human driver'}. Notice that a three-option scale rather than a forced choice scale (with no middle option \textit{`Not sure'}) was used because 1) humanness is more like a continuous rather than simple dichotomous variable; 2) using a three-option rating scale is a decent trade-off between an attempt to create an approximately continuous variable for the humanness (i.e., a rating scale with more options is better) and the convenience for passengers to ascribe humanness (i.e., a rating scale with fewer options is better).

Excepting all the quantitative ratings, the qualitative assessments, i.e., participants' mixed feelings, were also collected, given that the information contained in natural language texts may be able to predict human behaviour \cite{Rocklage_2021}. The lower-left corner of Fig.~\ref{fig2}D shows an example of one participant's mixed feelings about the past stage: `\begin{CJK*}{UTF8}{gkai}过红绿灯时停车较急促。\end{CJK*}' (The car stopped more quickly at traffic lights). In total, we got 68, 68, and 65 participants' data for the first, second, and third stages, respectively.

\subsubsection{Backbone: Signal detection theory \label{sec2.2.2}}

Signal detection theory (SDT) \cite{marcum1947statistical,tanner1954decision,green1966signal} is a general framework widely used by psychologists to describe decisions made under conditions of uncertainty. Here, we adopted the most common SDT framework, the equal variance SDT (EVSDT) model (which assumes that signal strength distributions are two Gaussian distributions with equal variances), as the backbone of our computational model, with the motivation to regard the perception system of the passenger as an information processing \cite{Broadbent_1958,Wickens_2012} system (Fig.~\ref{fig2}B). Thus, we could formulate passengers' ascription of humanness into detecting the signal from the noise, in which the stimulus ($E$ in Lewin's equation) from the human driver represents the signal, and that from the AI driver represents the noise.

To better introduce this information processing process, we take the input signal strength $\textit{SS}_k$, stimulus $E_k$ and output humanness rating behaviour $B_k$ of the observation $k$ as an example. We begin by calculating the point estimates of the EVSDT parameters. More specifically, using observations (excluding $k$) from passengers, we can compute hit rates $H_{1/2}$, $H_{2/3}$ and false alarm rates $F_{1/2}$ and $F_{2/3}$ under two criteria by (here, we hypothesised that the signal strength from the human driver was greater than that from the AI driver, i.e., \textit{hypothesis} 1 ($\textit{H}_{1}$), as depicted on the left of Fig.~\ref{fig2}B): 

\begin{align} 
\begin{split} 
H_{1/2} &= P (2 \cup 3 \,|\, \text{Human driver})\\
& =  \frac{\text{\#Observ. in which } B \text{ = } 2 \text{ or } 3, E \text{ = Human driver}}{\text{\#Human driver}}
\end{split}
\\%[1.5ex]
\begin{split} 
H_{2/3} &= P (3 \,|\, \text{Human driver})\\
& =  \frac{\text{\#Observ. in which } B \text{ = } 3, E \text{ = Human driver}}{\text{\#Human driver}}
\end{split}
\\[1.5ex]
\begin{split} 
F_{1/2} &= P (2 \cup 3 \,|\, \text{AI driver})\\
& =  \frac{\text{\#Observ. in which } B \text{ = } 2 \text{ or } 3, E \text{ = AI driver}}{\text{\#AI driver}}
\end{split}
\\
\begin{split} 
F_{2/3} &= P (3 \,|\, \text{AI driver})\\
& =  \frac{\text{\#Observ. in which } B \text{ = } 3, E \text{ = AI driver}}{\text{\#AI driver}}
\end{split}
\end{align}

Then response criteria $c_1$ and $c_2$ can be given by: 

\begin{align} 
c_1 = \left\{
\begin{aligned} 
&-\!\Phi^{-1}(H_{1/2}), & \text{if } E_k \text{ = Human driver}   \\
&-\!\Phi^{-1}(F_{1/2}), & \text{if } E_k \text{ = AI driver} 
\end{aligned}
\right.
\end{align}
\begin{align} 
c_2 = \left\{
\begin{aligned} 
&-\!\Phi^{-1}(H_{2/3}), & \text{if } E_k \text{ = Human driver}   \\
&-\!\Phi^{-1}(F_{2/3}), & \text{if } E_k \text{ = AI driver} 
\end{aligned}
\right.
\end{align}

where $\Phi^{-1}$ is the inverse cumulative normal distribution function, which converts the hit rate or false alarm rate into a $z$ score. Therefore, $B_k$ given $\textit{SS}_k$ and $E_k$ is:

\begin{align} 
B_k = M_k = \left\{
\begin{aligned} 
&1, & \text{if } \textit{SS}_k \leq c_1\\
&2, & \text{if } c_1 < \textit{SS}_k < c_2  \\
&3, & \text{if } \textit{SS}_k \geq c_2
\end{aligned}
\right.
\end{align}

Notice that in this example, $B_k$ equals the magnitude of $\textit{SS}_k$, i.e., $M_k$, which means $B$ and $M$ are positively correlated. An alternative hypothesis, \textit{hypothesis} 2 ($\textit{H}_{2}$), is that the signal strength from the AI driver was greater than that from the human driver, i.e., $B$ and $M$ are negatively correlated, as shown on the right of Fig.~\ref{fig2}B, and the above calculations can easily be adapted to $\textit{H}_{\textit{2}}$.

\subsubsection{Signal strength: Affective transition} \label{sec2.2.3}

\begin{figure*}[!h]
\center
\includegraphics[scale=0.668]{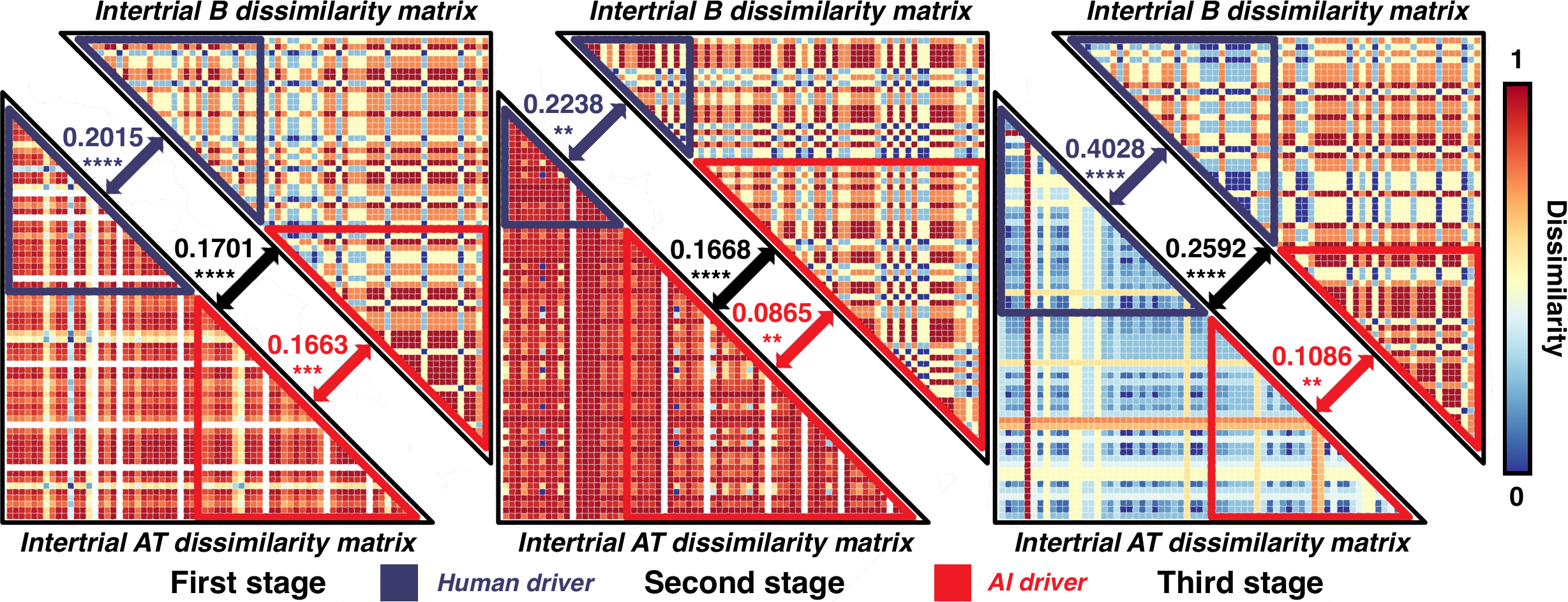}
\caption{Intertrial variability in affective transition (AT) was significantly and consistently correlated with intertrial variability in humanness rating behaviour ($B$) across three road stages and two conditions. Each triangular matrix of dissimilarity reflects intertrial variability in AT (derived from the distance between multidimensional scores of pre-study and post-stage emotions without transformation procedure) and $B$, respectively. All correlation scores are in Spearman \textit{rho} rank-order units (** $p < .01$, **** $p < .0001$), and related p-values were derived from one-tailed permutation tests (10,000 iterations).}
\label{fig3}
\end{figure*}

Further, to figure out how to represent the signal strength in SDT, we examined whether pre-study baseline emotions (including enjoyment, interest, surprise, fear, tension and satisfaction), post-stage emotions (same as baseline emotions) and safety and comfort scores are associated with passengers' humanness rating behaviour (i.e., $B$). None of these measures was consistently correlated with $B$ across three road stages (Appendix Table~\ref{tabA1}). Moreover, we also found that the raw scores of the measures mentioned above were not significantly different between the human and AI driver conditions (Appendix Fig.~\ref{figA1}), which ineluctably can not hold the role of signal strength to detect humanness across three stages. 

If not the above pre-study nor post-stage measures affect the passenger's humanness rating behaviour, then perhaps a dynamic change in emotions, i.e., affective transition (AT) between pre-study baseline emotions and corresponding post-stage emotions, holds the key. We tested this possibility by using representational similarity analysis (RSA) \cite{Kriegeskorte_2008}. RSA is a widely used framework for analysing common representational mapping between computational models, brain activity and behavioural data \cite{Kriegeskorte_2008,Popal_2019,lu2020neurora}, in which second-order isomorphism \cite{Shepard1970} (i.e., the match of dissimilarity matrices) is of the essence. By relating the representational geometry of affective transition to humanness rating behaviour, we found that intertrial variability in AT was significantly and consistently correlated with that in $B$ across three road stages and two conditions (Fig.~\ref{fig3}), indicating the potential of AT to play a role in passengers' ascription of humanness. Ergo, we employed AT, computed as the proximity between self-reported scores of pre-study and post-stage emotions, as the signal strength in SDT. That is to say, we leveraged passengers' AT to represent variable $P$ for investigating the specific and concrete form of Lewin's equation \cite{lewin1936} $B = f(P, E)$ in our case.

Continuing with the above example, i.e., observation $k$, we can compute $\textit{AT}_k$ as the distance between the pre-study baseline vector ($\bm{v}^{pre}_k$) and post-stage vector ($\bm{v}^{post}_k$) as follows:

\begin{align} 
\textit{AT}_k = \textit{SS}_k = z(dist(\bm{v}^{pre}_k, \bm{v}^{post}_k))
\end{align}

where $z$ denotes $z$-score normalisation, and $dist$ represents the distance measure, which could be absolute distance, one of the Anna Karenina distances \cite{finn2020idiosynchrony} (including mean distance, minimum distance and the product of the absolute and minimum distance), reversed Anna Karenina distance \cite{finn2020idiosynchrony} (i.e., maximum distance), Pearson distance, Euclidean distance, Mahalanobis distance, cosine distance, Manhattan distance, word mover's distance \cite{pmlr-v37-kusnerb15} or word rotator's distance \cite{Yokoi_2020}. Notice that the selection of specific distance measures was performed under the cross-validation procedure (Section~\ref{sec3.2}).

\subsubsection{Transformation: Leveraging pre-trained language models \label{sec2.2.4}}

Another key point to remember is that we transformed passengers' rating scores of emotions into corresponding language descriptions (together with their written mixed feelings) and leveraged PLMs to obtain the high-dimensional text representation to compute affective transition (Fig.~\ref{fig2}D).

The intuition of transformation is two-fold: Firstly, recent evidence from cognitive neuroscience has shown that, in addition to sensory-derived, embodied knowledge representation, there is another language-derived, non-sensory knowledge representation for concepts with sensory referents in the human brain \cite{striem2018neural,wang2020two,bi2021dual}, and PLMs hold great promise for simulating this type of knowledge coding system \cite{toneva2020combining,schrimpf2020neural,li2021cortical,goldstein2022shared}. Therefore, we may better represent passengers' emotional experiences by utilising PLMs to simulate the language-derived coding system in their brains. Secondly, PLMs have achieved unprecedented success in many natural language processing (NLP) tasks \cite{peters2018deep,devlin2019bert,brown2020language,Han_2021}. Incorporating the prior semantic knowledge in PLMs into our computational model might further boost the model performance. Thus, we could gain a better understanding of passengers' humanness rating behaviour from a data-driven perspective.

In this study, we tested 282 different PLMs, including 120 pre-trained word embeddings \cite{P18-2023,song2018directional} and 162 transformer-based \cite{vaswani2017attention} PLMs (such as ELECTRA \cite{clark2020electra} and T5 \cite{raffel2020exploring}), for encoding passenger's emotion scores and their written mixed feelings. Specifically, given the corresponding language descriptions $L_k$ of the observation $k$, we can use the following equations to describe the general \textbf{feature extraction} process of a multilayer transformer-based PLM:

\begin{equation}
\bm{H}^{0}_k=\bm{E}^{token}_k+\bm{E}^{seg}_k+\bm{E}^{pos}_k
\end{equation}

where $\bm{H}^{0}_k$ is the input representations constructed by summing the corresponding token embeddings ($\bm{E}^{token}_k$), segment embeddings ($\bm{E}^{seg}_k$), and position embeddings ($\bm{E}^{pos}_k$). Then, the hidden representations of $L_k$ at the $\alpha$-th layer of the $N$-layer PLM can be calculated as: 

\begin{equation}
\bm{H}^{\alpha}_k=transformer(\bm{H}^{\alpha-1}_k), \alpha \in [1, N]
\end{equation}

Empirically, we compute the average of hidden representations $\bm{H}^{avg}_k \in \mathbb{R}^{n \times d}$ from the first layer and last layer as the final extracted feature of $L_k$ \cite{DBLP:conf/emnlp/LiZHWYL20,su2021whitening}, where $n$ is the length of the $L_k$ and $d$ is the size of the transformer layer \footnote{For pre-trained word embeddings, we can directly get the corresponding hidden representations $\bm{H}_k$ for $L_k$ without the above procedure; nevertheless, we use the same symbol $\bm{H}^{avg}_k$ below for concise writing ($d$ denotes the number of dimensions for word embeddings in this case)}. Notice that we get the sentence-level representations via the above procedure. To get the document-level representations, we first need to get the sentence-level representations for six emotions and mixed feelings separately, and then conduct global average pooling over each matrix and stack these vectors vertically. Next, we conduct \textbf{global pooling} \cite{DBLP:journals/jmlr/CollobertWBKKK11} over $\bm{H}^{avg}_k$ to get the vector representation $\bm{v}_k \in \mathbb{R}^{d}$:

\begin{equation}
\bm{v}_k=pooling(\bm{H}^{avg}_k)
\end{equation}

where pooling operations could be max-, mean-, min-over-time operations or a combination of two or three of these operations. Finally, we further conduct \textbf{whitening transformation and dimensionality reduction} \cite{su2021whitening} to improve the representations obtained via the above procedure. Given a set of vector representations of $N$ observations $\{v_{i}\}^{N}_{i=1}$, we can compute its mean vector $\bm{\mu}$ and covariance matrix $\bm{\Sigma}$ as follows:

\begin{equation}
\bm{\mu}=\frac{1}{N}\sum^N_{i=1}\bm{v}_i
\end{equation}

\begin{equation}
\bm{\Sigma}=\frac{1}{N}\sum^N_{i=1}(\bm{v}_i-\bm{\mu})^T(\bm{v}_i-\bm{\mu})
\end{equation}

Then we conduct SVD decomposition \cite{Golub_1971} over $\bm{\Sigma}$ to get the related orthogonal matrix $\bm{U}$ and diagonal matrix $\bm{\Lambda}$. Let $\bm{W}=\sqrt{\bm{\Lambda}^{-1}}[:,\,:\kappa]$ ($\kappa \in [1, d/2]$ and $\kappa$ denotes the number of columns that need to be kept in $\bm{W}$), the \textbf{transformed vector} $\bm{v}_k$ can be given as: 

\begin{equation}
\bm{v}_k=(\bm{v}_k-\bm{\mu})\bm{W}
\end{equation}

The selection of different level representations, the specific pooling operations and the $\kappa$ value were performed under the cross-validation procedure (Section~\ref{sec3.2}).

\section{Results \label{sec3}}

\subsection{Results of the non-verbal variation of the Turing test \label{sec3.1}}

To examine whether the AI driver passed our non-verbal variation of the Turing test, we ran one-sample Wilcoxon tests on the average humanness rating scores (normalised to the range $[0, 1]$ for better illustration) across trials for each condition against the chance level of 0.5 (i.e., the expected value of random rating).
\begin{figure}[h]
\center
\includegraphics[scale=0.414]{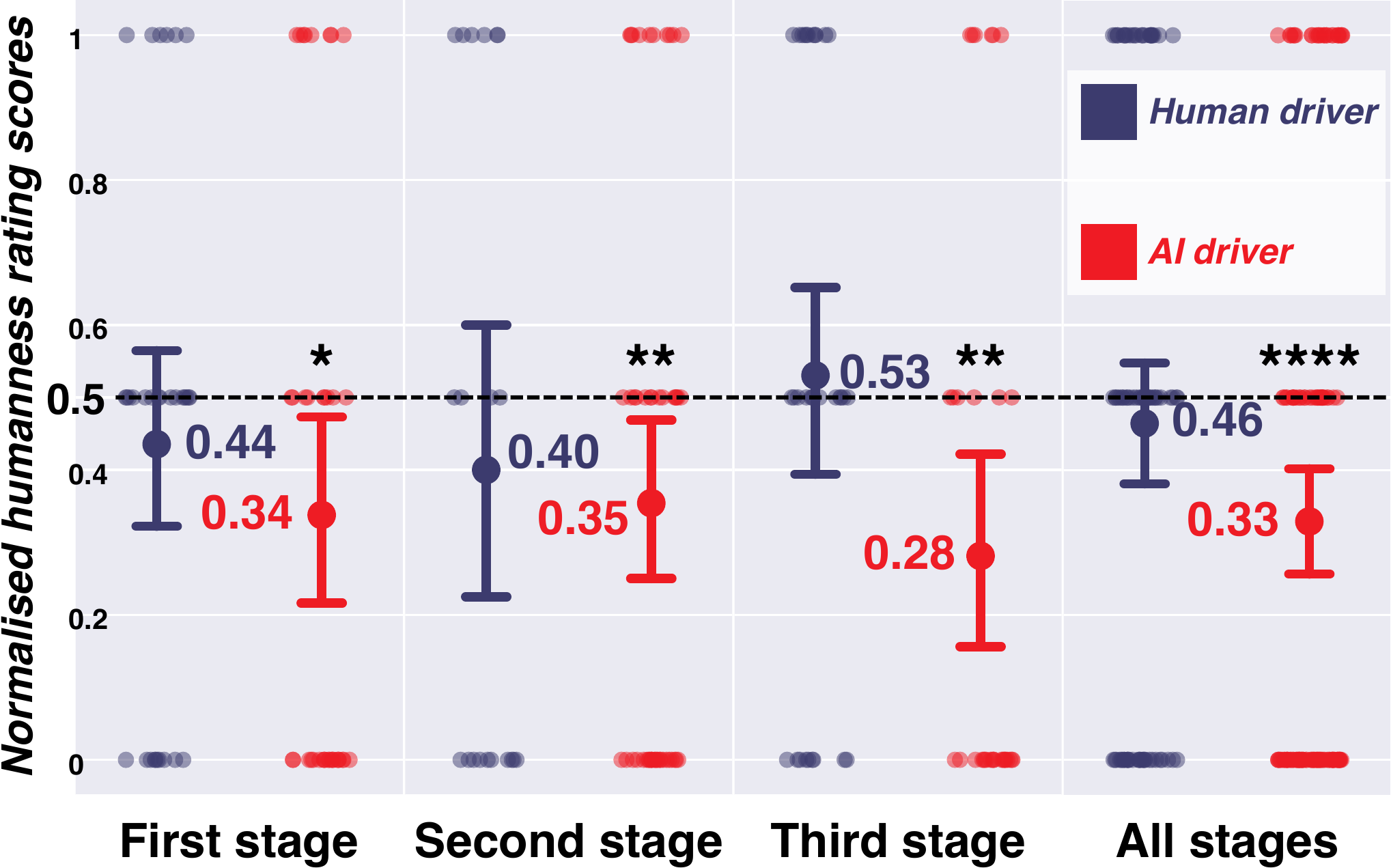}
\caption{The normalised humanness rating scores, their mean values and 95\% confidence intervals (CI) under different conditions. One-sample Wilcoxon tests confirmed that passengers' humanness ratings were significantly below the chance level (0.5 normalised humanness rating score) under the AI driver condition. In contrast, under the human driver condition, passengers' humanness ratings were at the chance level. * $p < .05$, ** $p < .01$, **** $p < .0001$.}
\label{fig4}
\end{figure}
As shown in Fig.~\ref{fig4}, when the AI driver controlled the AC, passengers' humanness ratings were significantly below the chance level across three separate road stages and all stages (first stage: CI = $[0.216, 0.473]$, $p$ = $0.012$; second stage: CI = $[0.240, 0.469]$, $p$ = $0.008$; third stage: CI = $[0.156, 0.422]$, $p$ = $0.003$; all stages: CI = $[0.256, 0.402]$, $p$ = $1.01\times10^{-5}$). While under the human driver condition, passengers' humanness ratings were at the chance level (first stage: CI = $[0.306, 0.565]$, $p$ = $0.322$; second stage: CI = $[0.225, 0.600]$, $p$ = $0.327$; third stage: CI = $[0.394, 0.667]$, $p$ = $0.662$; all stages: CI = $[0.381, 0.548]$, $p$ = $0.407$). The above results indicated that, on average, passengers could detect and discriminate between human and AI drivers. Thus, the AI driver did not pass our non-verbal variation of the Turing test.  

\subsection{Results of the computational models \label{sec3.2}}

\begin{table*}
\renewcommand{\arraystretch}{1.3}
\caption{Comparisons on the Outer Loop Cross-Validation of Nested-LOOCV with Baselines}
\label{tab2} 
\begin{subtable}[h]{\textwidth}
\centering
\caption{Evaluation results on the first stage.}
\label{tab21}
\begin{tabular}{p{13.1mm}<{\centering}||p{13.1mm}<{\centering}||p{13.1mm}<{\centering}||p{13.1mm}<{\centering}p{13.1mm}<{\centering}p{13.1mm}<{\centering}p{13.1mm}<{\centering}p{13.1mm}<{\centering}p{13.1mm}<{\centering}p{13.1mm}<{\centering}}
\hline\hline
Baselines & \multicolumn{1}{c}{\textit{AA}} & \multicolumn{1}{c}{$\textit{AA}_{pre}$} & $\textit{AA}_{post}$ & \textit{PA} & $\textit{PA}_{pre}$ & $\textit{PA}_{post}$ & \textit{NA} & $\textit{NA}_{pre}$ & $\textit{NA}_{post}$ \\ \hline\hline
MLR & \multicolumn{1}{c}{-0.1844} & \multicolumn{1}{c}{0.1312} & 0.1283 & 0.0988 & 0.1761 & -0.0082 & -0.0453 & 0.0390 & 0.0744 \\
KNN & \multicolumn{1}{c}{0.1431} & \multicolumn{1}{c}{0.0543} & 0.1753 & 0.4755**** & 0.2370* & -0.0669 & 0.0870 & -0.1078 & 0.1129 \\
SVC & \multicolumn{1}{c}{-0.1039} & \multicolumn{1}{c}{-0.1027} & -0.0268 & 0.1704 & 0.0431 & -0.0932 & 0.0780 & 0.0340 & -0.0578 \\
RF & \multicolumn{1}{c}{-0.0654} & \multicolumn{1}{c}{0.1239} & -0.0122 & 0.1125 & 0.1245 & -0.2744 & 0.0688 & 0.0586 & 0.1301 \\
XGBoost & \multicolumn{1}{c}{0.1794} & \multicolumn{1}{c}{0.4125***} & 0.0537 & 0.2188* & 0.0754 & 0.0430 & 0.1013 & 0.1508 & 0.1321 \\
MLP & \multicolumn{1}{c}{0.2185*} & \multicolumn{1}{c}{0.3211**} & -0.1391 & -0.0759 & 0.1083 & 0.0953 & 0.0448 & -0.1041 & 0.0342 \\
\hline\hline
Baselines & \textit{None} & \textbf{SDT-AT} & \textit{AA+MF} & \textit{AA} & \textbf{\textit{PA+MF}} & \textit{PA} & \textit{NA+MF} & \textit{NA} & \textit{MF} \\ \hline\hline
Random      & 0.0029 & Original & -0.3985 & -0.3552 & -0.2580 & 0.1738 & -0.3397 & 0.0828 & 0.0990 \\
Probability & -0.0060 & PLM-wv & 0.4511*** & 0.4152*** & 0.4092*** & 0.3939*** & 0.4064*** & 0.1359 & 0.3030** \\
Detective   & 0.1491 & \textbf{PLM-tf} & 0.4113*** & 0.4639**** & \textbf{0.4768****} & 0.3939*** & 0.3484** & 0.1842 & 0.3738** \\ \hline\hline
\end{tabular}
\end{subtable}
\\[5.05pt]
\begin{subtable}[h]{\textwidth}
\centering
\caption{Evaluation results on the second stage.}
\label{tab22}
\begin{tabular}{p{13.1mm}<{\centering}||p{13.1mm}<{\centering}||p{13.1mm}<{\centering}||p{13.1mm}<{\centering}p{13.1mm}<{\centering}p{13.1mm}<{\centering}p{13.1mm}<{\centering}p{13.1mm}<{\centering}p{13.1mm}<{\centering}p{13.1mm}<{\centering}}
\hline\hline
Baselines & \multicolumn{1}{c}{\textit{AA}} & \multicolumn{1}{c}{$\textit{AA}_{pre}$} & $\textit{AA}_{post}$ & \textit{PA} & $\textit{PA}_{pre}$ & $\textit{PA}_{post}$ & \textit{NA} & $\textit{NA}_{pre}$ & $\textit{NA}_{post}$ \\ \hline\hline
MLR & \multicolumn{1}{c}{0.2752*} & \multicolumn{1}{c}{0.1524} & -0.2298 & 0.1539 & 0.2095* & -0.1659 & 0.0205 & 0.1947 & -0.1728 \\
KNN & \multicolumn{1}{c}{0.2046*} & \multicolumn{1}{c}{0.3069**} & -0.3189 & 0.1436 & 0.1297 & -0.3123 & -0.2696 & -0.1486 & -0.1639 \\
SVC & \multicolumn{1}{c}{0.1061} & \multicolumn{1}{c}{0.0945} & -0.1743 & 0.1270 & -0.0558 & -0.0776 & 0.0161 & 0.0541 & 0.0997 \\
RF & \multicolumn{1}{c}{0.0416} & \multicolumn{1}{c}{0.3126**} & -0.1799 & 0.2379* & 0.2588* & -0.2196 & 0.0573 & 0.2087* & -0.3861 \\
XGBoost & \multicolumn{1}{c}{0.0835} & \multicolumn{1}{c}{0.2839**} & -0.2254 & 0.1895 & 0.3613** & -0.1368 & -0.0965 & -0.2473 & -0.1788 \\
MLP & \multicolumn{1}{c}{0.1986} & \multicolumn{1}{c}{0.1981} & -0.3661 & 0.1302 & 0.3687** & -0.1213 & -0.0608 & -0.3048 & -0.3838 \\ \hline\hline
Baselines & \textit{None} & \textbf{SDT-AT} & \textit{AA+MF} & \textit{AA} & \textit{PA+MF} & \textbf{\textit{PA}} & \textit{NA+MF} & \textit{NA} & \textit{MF} \\ \hline\hline
Random      & 0.0010 & Original & 0.1750 & 0.2409* & 0.1539 & 0.1912 & 0.1865 & -0.0105 & 0.1824 \\
Probability & -0.0017 & PLM-wv & 0.4569**** & 0.4195*** & 0.4402*** & 0.4635**** & 0.3167** & 0.1703 & 0.4276*** \\
Detective   & 0.0394 & \textbf{PLM-tf} & 0.4375*** & 0.4173*** & 0.4545**** & \textbf{0.4739****} & 0.3528** & 0.2636* & 0.3578** \\ \hline\hline
\end{tabular}
\end{subtable}
\\[5.05pt]
\begin{subtable}[h]{\textwidth}
\centering
\caption{Evaluation results on the third stage.}
\label{tab23}
\begin{tabular}{p{13.1mm}<{\centering}||p{13.1mm}<{\centering}||p{13.1mm}<{\centering}||p{13.1mm}<{\centering}p{13.1mm}<{\centering}p{13.1mm}<{\centering}p{13.1mm}<{\centering}p{13.1mm}<{\centering}p{13.1mm}<{\centering}p{13.1mm}<{\centering}}
\hline\hline
Baselines & \multicolumn{1}{c}{\textit{AA}} & \multicolumn{1}{c}{$\textit{AA}_{pre}$} & $\textit{AA}_{post}$ & \textit{PA} & $\textit{PA}_{pre}$ & $\textit{PA}_{post}$ & \textit{NA} & $\textit{NA}_{pre}$ & $\textit{NA}_{post}$ \\ \hline\hline
MLR & \multicolumn{1}{c}{0.2154*} & \multicolumn{1}{c}{0.3482**} & 0.2852* & 0.0593 & -0.0535 & 0.0076 & 0.3994*** & 0.3294** & 0.3954*** \\
KNN & \multicolumn{1}{c}{0.1782} & \multicolumn{1}{c}{0.4317***} & 0.2630* & 0.0885 & 0.1510 & 0.1899 & 0.3998*** & 0.4161*** & 0.3301** \\
SVC & \multicolumn{1}{c}{0.1425} & \multicolumn{1}{c}{0.3438**} & 0.2218* & -0.0157 & -0.0608 & 0.1165 & 0.1932 & 0.1456 & 0.3215** \\
RF & \multicolumn{1}{c}{0.1180} & \multicolumn{1}{c}{0.3615**} & 0.0360 & 0.0654 & 0.1642 & 0.0294 & 0.3397** & 0.2815* & 0.3244** \\
XGBoost & \multicolumn{1}{c}{0.2186*} & \multicolumn{1}{c}{0.3625**} & 0.1942 & 0.0674 & 0.1525 & 0.1175 & 0.3339** & 0.4016*** & 0.2987** \\
MLP & \multicolumn{1}{c}{0.1302} & \multicolumn{1}{c}{0.2144*} & 0.2740* & 0.0347 & 0.0722 & 0.2187* & 0.3674** & 0.3126** & 0.2512* \\ \hline\hline
Baselines & \textit{None} & \textbf{SDT-AT} & \textit{AA+MF} & \textit{AA} & \textit{PA+MF} & \textit{PA} & \textit{NA+MF} & \textit{NA} & \textit{\textbf{MF}} \\ \hline\hline
Random      & 0.0001 & Original & 0.1490 & 0.2019 & 0.1978 & -0.0258 &  0.4037*** & 0.4245*** & 0.1104 \\
Probability & -0.0021 & \textbf{PLM-wv} & 0.4861**** & 0.4556*** & 0.4624*** & 0.4322*** & 0.4419*** & 0.4256*** & \textbf{0.5615****} \\
Detective   & 0.3168** & PLM-tf & 0.4807**** & 0.4974**** & 0.4654**** & 0.4570*** & 0.4769**** & 0.4429*** & 0.5422**** \\ \hline\hline
\end{tabular}
\end{subtable}
\\[5.05pt]
\begin{subtable}[h]{\textwidth}
\centering
\caption{Evaluation results on all stages.}
\label{tab24}
\begin{tabular}{p{13.1mm}<{\centering}||p{13.1mm}<{\centering}||p{13.1mm}<{\centering}||p{13.1mm}<{\centering}p{13.1mm}<{\centering}p{13.1mm}<{\centering}p{13.1mm}<{\centering}p{13.1mm}<{\centering}p{13.1mm}<{\centering}p{13.1mm}<{\centering}}
\hline\hline
Baselines & \multicolumn{1}{c}{\textit{AA}} & \multicolumn{1}{c}{$\textit{AA}_{pre}$} & $\textit{AA}_{post}$ & \textit{PA} & $\textit{PA}_{pre}$ & $\textit{PA}_{post}$ & \textit{NA} & $\textit{NA}_{pre}$ & $\textit{NA}_{post}$ \\ \hline\hline
MLR & \multicolumn{1}{c}{0.0573} & \multicolumn{1}{c}{0.1516*} & 0.0749 & 0.0543 & 0.1264* & 0.0988 & 0.0931 & 0.1160 & 0.0520 \\
KNN & \multicolumn{1}{c}{0.0992} & \multicolumn{1}{c}{0.1521*} & 0.1198* & 0.0419 & 0.0144 & 0.1216* & 0.1116 & 0.1422* & -0.0497 \\
SVC & \multicolumn{1}{c}{0.0854} & \multicolumn{1}{c}{0.0755} & 0.1457* & 0.0414 & 0.0991 & 0.0688 & 0.0467 & 0.0676 & 0.0038 \\
RF & \multicolumn{1}{c}{0.0505} & \multicolumn{1}{c}{0.1308*} & 0.0292 & 0.1491* & 0.0457 & -0.0001 & 0.0117 & 0.0500 & 0.1426* \\
XGBoost & \multicolumn{1}{c}{0.1411*} & \multicolumn{1}{c}{0.2586***} & 0.0198 & 0.1254* & 0.1157 & 0.0044 & 0.2176** & 0.1969** & 0.1357* \\
MLP & \multicolumn{1}{c}{0.0952} & \multicolumn{1}{c}{0.1949**} & 0.0701 & 0.1349* & 0.0540 & 0.0830 & 0.2037** & 0.2078** & 0.0842 \\ \hline\hline
Baselines & \textit{None} & \textbf{SDT-AT} & \textit{AA+MF} & \textit{AA} & \textit{PA+MF} & \textit{PA} & \textit{NA+MF} & \textit{NA} & \textbf{\textit{MF}} \\ \hline\hline
Random      & 0.0013 & Original & 0.1850** & 0.1816** & 0.0326 & 0.1416* & -0.1204 & 0.1685** & 0.0570 \\
Probability & -0.0006 & \textbf{PLM-wv} & 0.2704*** & 0.2452*** & 0.2447*** & 0.2331*** & 0.2866**** & 0.1871** & \textbf{0.5093****} \\
Detective   & 0.1764** & PLM-tf & 0.2837**** & 0.2879**** & 0.2734**** & 0.2878**** & 0.4178**** & 0.2004** & 0.4641**** \\ \hline\hline
\end{tabular}
\end{subtable}
\\[5.05pt]
\small All the evaluation results are in Spearman \textit{rho} rank-order correlation units (* $p < .05$, ** $p < .01$, *** $p < .001$, **** $p < .0001$). The winning models and the related results are bold. All the $p$-values were based on one-tailed tests of significance. `AA' for all affect, `PA' for positive affect, `NA' for negative affect and `MF' for mixed feelings; `pre' for pre-study baseline and `post' for post-stage.
\end{table*}

We trained and evaluated the computational models under the nested leave-one-out cross-validation (nested-LOOCV) procedure \cite{Hastie_2001}. We compared our models with machine learning baselines \footnote{Due to the high computational load, we trained and evaluated machine learning baselines under the nested cross-validation with ten folds in the outer loop and five folds in the inner loop.}: MLR, the multi-class logistic regression classifier; KNN, the nearest neighbour classifier; SVC, the support vector machine; RF, the random forest classifier; XGBoost, the decision tree-based ensemble classifier that uses a gradient boosting framework; MLP, the multilayer perceptron classifier; and naive baselines: Random, which posits that the passenger's humanness rating behaviour is generated at random with equal probability; Probability, which posits that the passenger's humanness rating behaviour is drawn at random from the population of history ratings; Detective, which posits that the passenger could discern the difference between the human and AI drivers and thus make the correct guess all the time. 

Within our proposed SDT-AT framework, we tested the following models: Original, in which AT was derived directly from a distance between multidimensional scores of pre-study and post-stage emotions without transformation with PLM; PLM-wv, in which pre-trained word embeddings would transform passenger's emotion scores or mixed feelings; PLM-tf, in which transformer-based PLM would transform passenger's emotion scores or mixed feelings. For the above SDT-AT models, we computed AT based on different data components: positive affect (PA, including enjoyment, interest, surprise and satisfaction), negative affect (NA, including fear and tension), all affect (AA, including PA and NA), mixed feelings (MF) \footnote{For the Original model, MF would be 1 if the passenger had written mixed feelings or 0 if the passenger did not have written mixed feelings at a given trial.} or a combination between MF and other data components. For machine learning baselines, we tested different model inputs: AA, PA or NA of pre-study baselines, post-stage or a combination of the above two. We used Spearman's rank correlation score (\textit{rho}) as the evaluation metric and selected hyperparameters with the highest \textit{rho} score in the inner loop cross-validation of the nested-LOOCV.

The performance of different computational models is shown in Table~\ref{tab2} \footnote{For the convenience of the display, we only show the results of PLMs with the highest \textit{rho} score from the outer loop cross-validation of nested-LOOCV.}. Based on Lewin's equation, our proposed SDT-AT models provided superior within- (Table~\ref{tab2}a-c) and cross-stage performance (Table~\ref{tab24}) than all other baselines, demonstrating the overall effectiveness of these models. Moreover, the declining performance of the Original model indicated the significance of transformation with PLM in computing AT. Specifically, PLM-tf (PA+MF) and PLM-tf (PA) outperformed all other models with \textit{rho} scores of 0.4768 ($p$ = $3.94\times10^{-5}$) and 0.4739 ($p$ = $4.46\times10^{-5}$) on the first and second stages (Table~\ref{tab2}a-b, see further analysis in Section~\ref{sec4.2}), respectively. Furthermore, PLM-wv (MF) surpassed all the other competing models with \textit{rho} scores of 0.5615 ($p$ = $1.14\times10^{-6}$) and 0.5093 ($p < 1.0\times10^{-13}$) on the third and all stages (Table~\ref{tab2}c-d, see further analysis in Section~\ref{sec4.3}), respectively. 

\begin{figure}[!t]
\center
\includegraphics[scale=0.56]{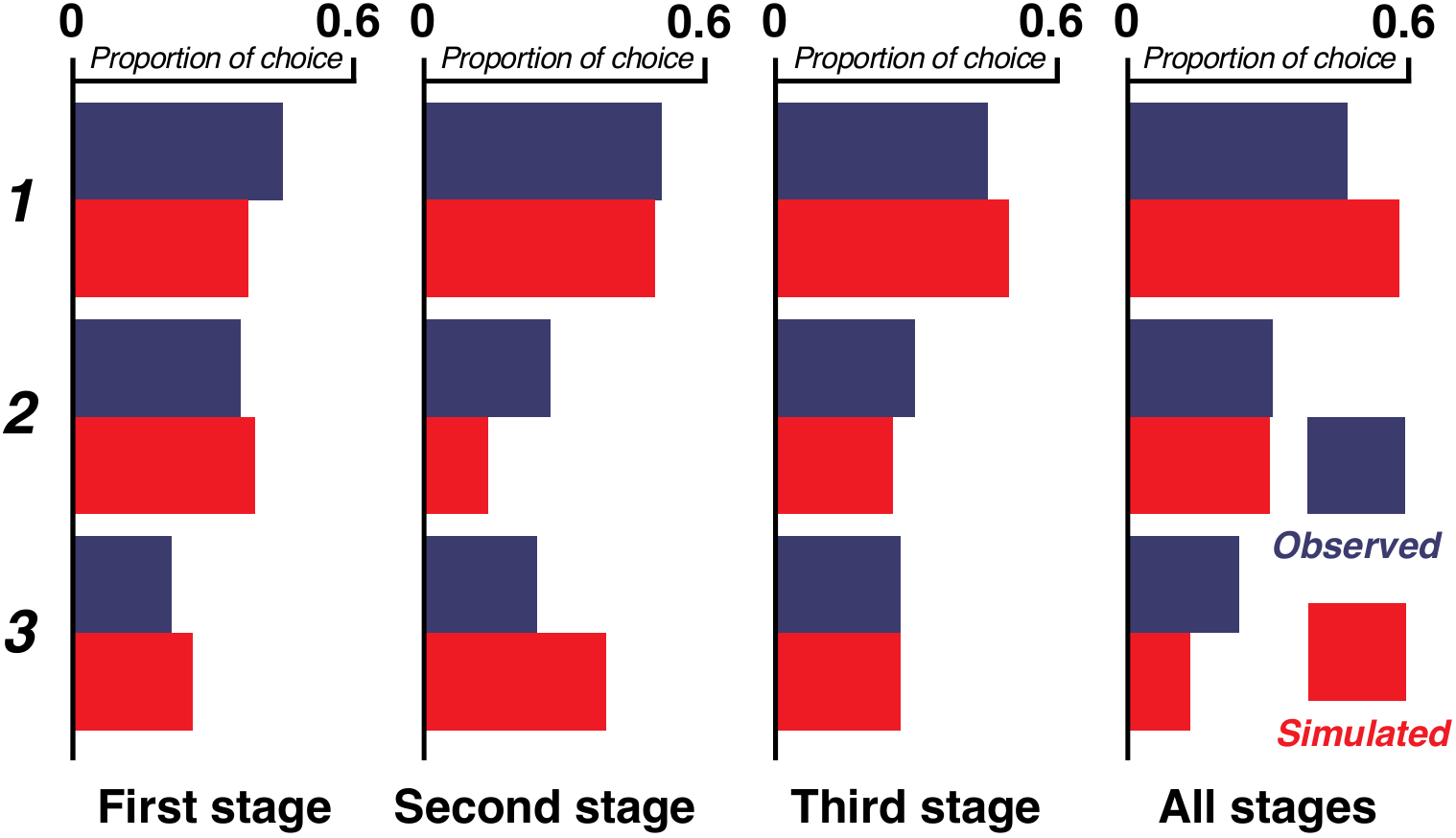}
\caption{Comparisons of the proportion of humanness rating scores to a variation of a Turing test question between empirical observations (blue) and model simulations (red). 1 for \textit{`AI driver'}; 2 for \textit{`Not sure'}; 3 for \textit{`Human driver'}.}
\label{fig5}
\end{figure}

\begin{figure}[!t]
\center
\includegraphics[scale=0.44]{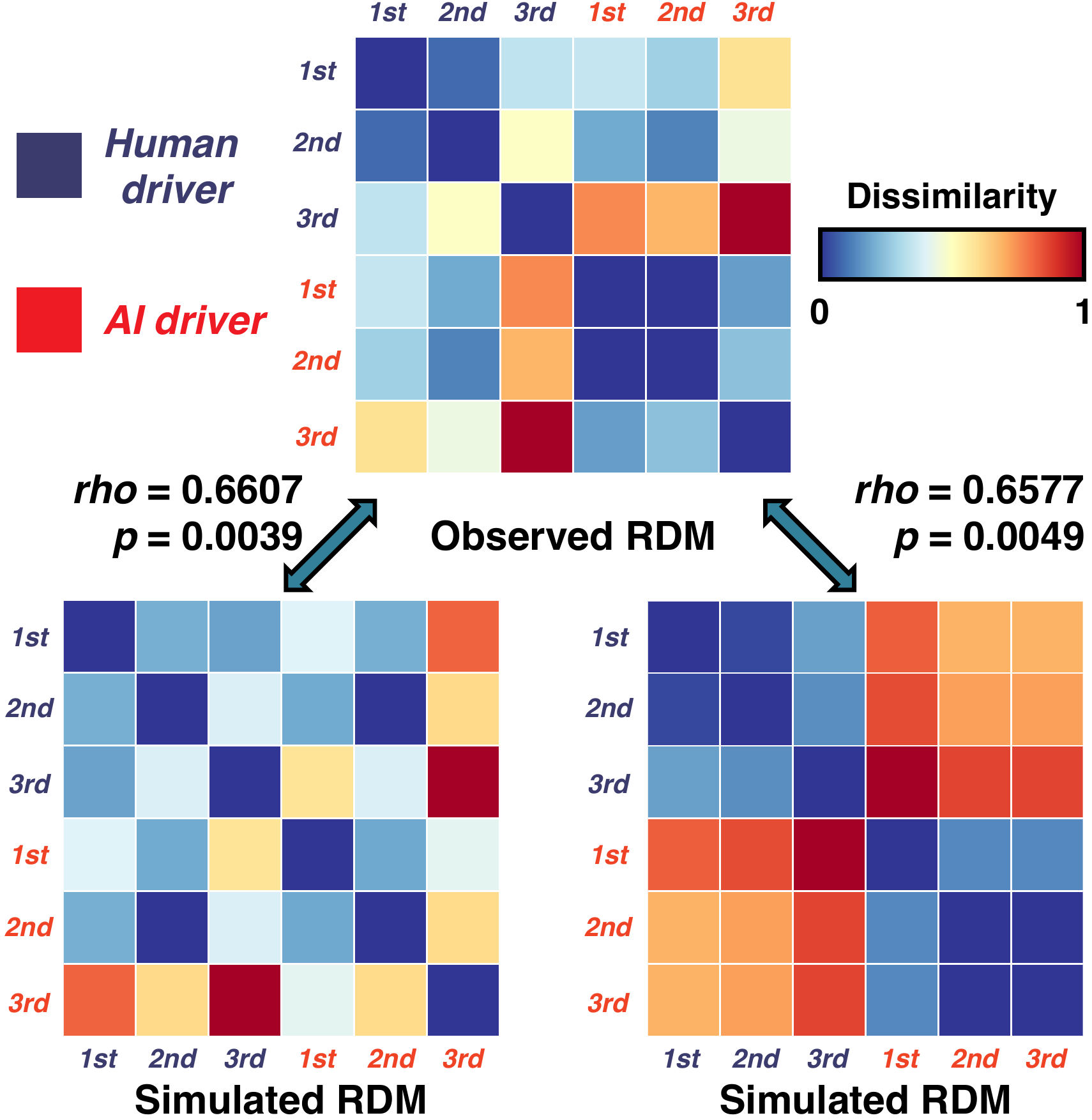}
\caption{Representational similarity between empirically observed humanness rating scores to a variation of a Turing test question (top representational dissimilarity matrix, RDM) and model simulations (bottom left RDM, derived from a combination of within-stage model simulations; bottom right RDM, derived from cross-stage model simulations) averaged over all trials. The blue and red colours denote stimuli, i.e., human and AI drivers, respectively. `1st' for the first stage; `2nd' for the second stage; `3rd' for the third stage. Both correlation scores are in Spearman \textit{rho} rank-order units, and corresponding p-values were derived from one-tailed permutation tests (10,000 iterations).}
\label{fig6}
\end{figure}

We also conducted the model simulations to verify whether our proposed winning computational models could replicate the passenger's humanness rating behaviour. Fig.~\ref{fig5} shows that our computational model accurately captured the passenger's humanness rating behaviour patterns in the non-verbal variation of the Turing test. Further, by using representational similarity analysis (RSA) \cite{Kriegeskorte_2008}, we directly compared the representational geometry of empirically observed humanness rating behaviour with those of model simulations. As shown in Fig.~\ref{fig6}, we found that representational dissimilarity matrices (RDMs) of model simulations were highly correlated with the RDM of empirically observed humanness rating behaviour (within-stage: $\textit{rho = } 0.6607$, $p$ = $0.0039$; cross-stage: $\textit{rho = } 0.6577$, $p$ = $0.0049$), suggesting that our computational model exhibited the same humanness rating behaviour pattern as passengers did. Altogether, these results permit us to use our computational models in further elucidating the implications that radiate from passengers' ascription of humanness in the non-verbal variation of the Turing test (see Section~\ref{sec4}).

\section{Analysis \label{sec4}}

\subsection{Analysis of relatedness between the humanness rating and the magnitude of affective transition \label{sec4.1}}

In computational modelling, we incorporated two competing hypotheses ($\textit{H}_{1}$ and $\textit{H}_{2}$, see Section~\ref{sec2.2.2} and Fig.~\ref{fig2}B) about the relatedness between humanness rating behaviour and the magnitude of affective transition into our proposed SDT-AT models, respectively. Then, we selected the winning model with the highest \textit{rho} score in the outer loop cross-validation of the nested-LOOCV, as reported in Table~\ref{tab2}. To reveal which hypothesis holds, we compared the passenger's humanness rating to the magnitude of AT derived from our winning models. 

\begin{figure}[h]
\center
\includegraphics[scale=0.68]{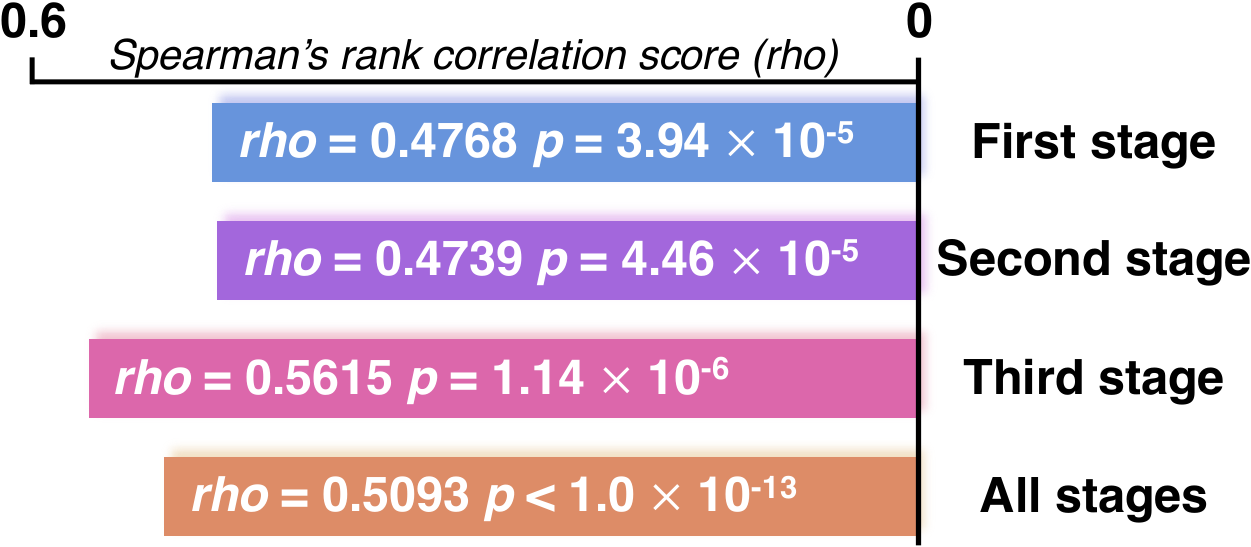}
\caption{Bar chart for the Spearman's rank correlation scores between the humanness rating (\textit{AI driver} 1, \textit{Not sure} 2, \textit{Human driver} 3) and the magnitude of affective transition (\textit{low magnitude} 1, \textit{medium magnitude} 2, and \textit{high magnitude} 3). Each trial's magnitude of affective transition was obtained using the nested-LOOCV procedure with the predictive model trained by the remaining $N - 1$ samples, excluding one to-be-predicted sample. All the $p$-values were based on two-tailed tests of significance. For the first and the second stage, $N\text{ = }68$; for the third stage, $N\text{ = }65$; for all stages, $N\text{ = }201$.}
\label{fig7}
\end{figure}

In favour of $\textit{H}_{1}$, we observed strong positive within- and cross-stage associations between the humanness rating and the magnitude of AT (first stage: $\textit{rho = } 0.4768$, $p$ = $3.94\times10^{-5}$; second stage: $\textit{rho = } 0.4739$, $p$ = $4.46\times10^{-5}$; third stage: $\textit{rho = } 0.5615$, $p$ = $1.14\times10^{-6}$; all stages: $\textit{rho = } 0.5093$, $p < 1.0\times10^{-13}$, see Fig.~\ref{fig7}), such that the ascription of humanness would increase with the greater affective transition. The above analysis suggested that AT, posited as a crucial part of passengers' ride experience in our model, may play an important role in their ascription of humanness.

\subsection{Analysis of the direction for AT on the starting two stages\label{sec4.2}}

Our proposed SDT-AT models in which AT was derived (or partly derived) from the positive affect (PA) dominated comparisons on the first and second stages (see Table~\ref{tab2}a-b). However, we did not know how the passenger's PA changed under two conditions during the starting two stages since AT is just a scalar quantity with no direction. It might be the case that the passenger's PA would greatly or moderately increase or decrease under the human driver condition while moderately or slightly increase or decrease under the AI driver condition, given that the previous analysis (Section~\ref{sec4.1}) showed that the signal strength (i.e., AT) from the human driver was greater than that from the AI driver. To investigate this further, we examined mean changes in PA (calculated as post-stage minus pre-study PA summary scores of enjoyment, interest, surprise and satisfaction) during the first and second stages, respectively. As presented in Table~\ref{tab3}, passengers showed significant or marginal significant increases in PA under the human driver condition (first stage: $\Delta\textit{M = } 0.742$, $p$\text{ = }$0.046$; second stage: $\Delta\textit{M = } 0.500$, $p$\text{ = }$0.065$), while passengers showed decreases in PA under the AI driver condition, though insignificantly (first stage: $\Delta\textit{M = --}\,0.622$, $p$\text{ = }$0.218$; second stage: $\Delta\textit{M = --}\,0.375$, $p$\text{ = }$0.223$). Our analysis indicates that enhancing positive affect may be the essence of the human-like ride experience during the starting two stages. Because the greater the affective transition along with this enhancement, the higher the passenger's humanness rating will be.

\begin{table}[t]
\renewcommand{\arraystretch}{1.3}
\centering
\caption{Mean Changes in Positive Affect During the First and Second Stages}  
\label{tab3} 
\begin{tabular}{c||cccc}
\hline\hline
Conditions & $\Delta$\textit{M} & \textit{SD} & $z$ & $p$ \\ \hline\hline
\multicolumn{1}{l||}{\textit{First stage}} & \multicolumn{1}{l}{} & \multicolumn{1}{l}{} & \multicolumn{1}{l}{} & \multicolumn{1}{l}{} \\ 
\multicolumn{1}{l||}{\quad \ \ \textbf{Human driver}} & \textbf{0.742} & \textbf{2.627} & \textbf{1.68} & \textbf{0.046} \\
\multicolumn{1}{l||}{\quad \ \ AI driver} & -0.622 & 2.803 & -0.78 & 0.218 \\ \hline\hline
\multicolumn{1}{l||}{\textit{Second stage}} & \multicolumn{1}{l}{} & \multicolumn{1}{l}{} & \multicolumn{1}{l}{} & \multicolumn{1}{l}{} \\
\multicolumn{1}{l||}{\quad \ \ \textbf{Human driver}} & \textbf{0.500} & \textbf{1.396} & \textbf{1.51} & \textbf{0.065} \\
\multicolumn{1}{l||}{\quad \ \ AI driver} & -0.375 & 2.983 & -0.76 & 0.223 \\ \hline\hline
\end{tabular}
\\[6pt]
\justifying
\noindent
\small Significant ($p < .05$) and marginal significant ($p < .1$) effects are bold. All the $p$-values were derived from one-tailed Wilcoxon signed-rank tests.
\end{table}

\subsection{Word cloud analysis of mixed feelings\label{sec4.3}}

Given our proposed SDT-AT models in which AT 
\begin{figure}[h]
\center
\includegraphics[scale=0.40875]{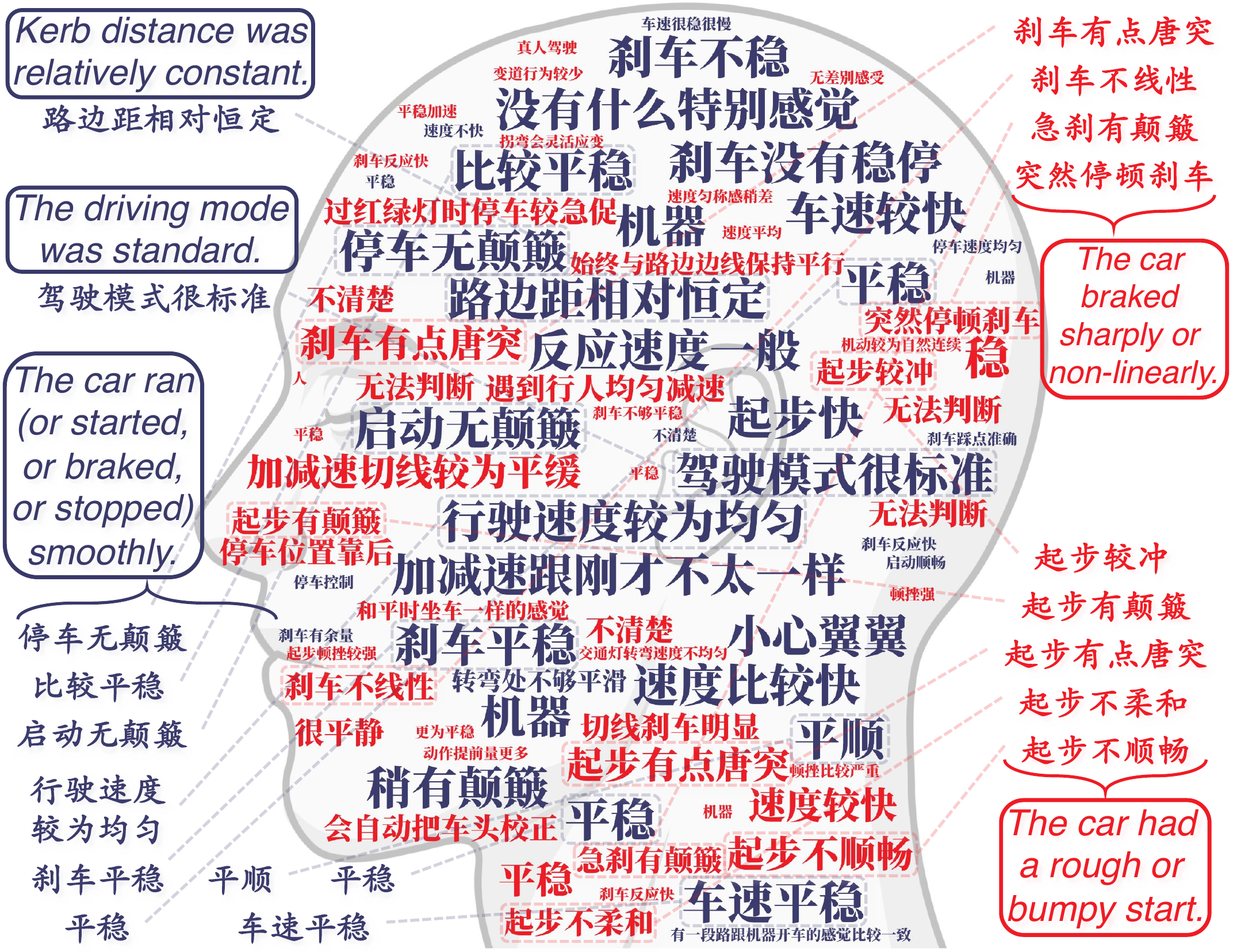}
\caption{Word cloud displaying mixed feelings (MF) from all stages. The blue and red colours denote stimuli, i.e., human and AI drivers, respectively. The size of each MF item is proportional (positively for the human driver condition, negatively for the AI driver condition) to the related $z$-scored transition from cross-stage model simulations. The defining MF items (i.e., those with larger sizes) in predicting passengers' humanness rating behaviour consistent with the stimulus (i.e., \textit{`AI driver'}, 1 and \textit{`Human driver'}, 3) were annotated. Among these defining items, those with similar semantics were clustered and translated to corresponding English. Figure created with BioRender.com and weiciyun.com.}
\label{fig8}
\end{figure}
was obtained from the mixed feelings (MF) yielded the best performance on the third and all stages (see Table~\ref{tab2}c-d), we further conducted word cloud analysis to compare the difference of MF induced by the human and AI driver. As shown in Fig.~\ref{fig8}, the word cloud highlights the salient MF items (i.e., those with larger sizes) under each condition, with the size of each MF item proportioning (positively or negatively for the human or AI driver condition) its $z$-scored AT from cross-stage model simulations. Specifically, under the human driver condition, the defining MF items in predicting passengers' highest humanness rating, \textit{`Human driver'} (3), were clustered (based on their semantics) as follows: `Kerb distance was relatively constant.'; `The driving mode was standard.'; `The car ran (or started, or braked or stopped) smoothly.' While under the AI driver condition, the defining MF items in predicting passengers' lowest humanness rating, \textit{`AI driver'} (1), were clustered as follows: `The car braked sharply or non-linearly.'; `The car had a rough or bumpy start.' The above comparison vividly showed the difference in the passenger's subjective ride experience between the two conditions and illustrates details of what needs to be improved for current automated driving to offer a human-like ride experience for the passenger and thus increase the social acceptance of ACs. 

\section{Discussion \label{sec5}}

\subsection{Contributions and implications}

As autonomous cars are increasing on our roads, the human role gradually shifts from active drivers to passive passengers. Meanwhile, a growing body of literature \cite{al2001framework,al2003toward,gu2017human,hecker2019learning,sun2020exploring} highlights that the acceptance of the AC will increase if it drives in a stereotypical human manner. Nevertheless, very little research has been devoted to investigating the humanness of the AC from the perspective of passive passengers. Herein, in the present study, for the first time, we examined whether the current SAE Level 4 AC, i.e., AC with the WeRide ONE \cite{Weride} as its self-driving algorithm, could create a human-like ride experience for passengers in a real-road scenario and hence pass the non-verbal variation of the Turing test from the perspective of passive passengers.

Our results showed that human passengers might be sensitive to the human-like ride experience, as indicated by the higher humanness rating in our non-verbal variation of the Turing test for the human driver condition relative to the AI driver condition. When the AI driver controlled the AC, results showed that passengers' humanness ratings were below the chance level, indicating that the WeRide ONE did not pass our variation of a non-verbal Turing test because human passengers could successfully detect the AI driver based on their subjective ride experience (Fig.~\ref{fig4}). Nonetheless, we also noticed that the WeRide ONE could successfully trick human passengers in some trials, revealing the promising fact that some self-driving algorithms, like the WeRide ONE, are beginning to learn and imitate human behaviour in a convincing manner.

As the literature suggests \cite{Waytz_2014}, even the best technology, such as a vehicle that drives itself, is of little use if the user does not accept it. Consequently, given the key role that human likeness played in improving the passengers' acceptance towards ACs, we investigated further why passengers could discern the AI driver in most trials and not in others in our non-verbal variation of the Turing test. Specifically, on the basis of Lewin's field theory \cite{lewin1936}, we advanced a computational model combining signal detection theory (SDT) with pre-trained language models (PLMs) to predict passengers' humanness rating behaviour. We employed affective transition (AT), computed as the proximity between rating vectors of pre-study and post-stage emotions transformed by PLM, as the signal strength in our SDT models. The results showed that our SDT-AT models could adequately predict passengers' humanness rating behaviour in the non-verbal variation of the Turing test (Table~\ref{tab2}, Fig.~\ref{fig5} and Fig.~\ref{fig6}), the implications of which are as follows. 

First, our proposed computational model is a concrete application of Lewin's field theory, in which we replaced the variables in Lewin's equation with the specific situational and personal characteristics of the passenger (e.g., $B$ with the humanness rating behaviour, $P$ with AT and $E$ with the stimulus). The practical success of basing the computational modelling on Lewin's seemingly abstract and theoretical field theory speaks directly to his famous maxim that `there is nothing as practical as a good theory' \cite{lewin1943psychology}. Second, our proposed models not only achieved superior within-stage performance than all other baselines (Table~\ref{tab2}a-c) but also showed superiority in cross-stage performance (Table~\ref{tab24}). Together with the agreement between model simulations and empirical observations (Fig.~\ref{fig5} and Fig.~\ref{fig6}), our results indicate that we may succeed in discovering the general law $B = f(P, E)$ which is valid for the dynamic structure of the passenger's psychological field (i.e., $(P,E)$). Finally, these results also demonstrate the possibility and feasibility of using NLP techniques, such as PLMs, as adjuncts to the interaction between social cognition and artificial intelligence to guide theorising and the generation of conceptual insights \cite{jackson2022text,tallon2022topological}.

Overall, conducting affective computing in this novel way enable us to discover the latent relatedness between AT and the passenger's humanness rating behaviour. Importantly, we offer the first insights into what renders passengers' subjective ride experience truly human-like for future automated driving: the passengers' ascription of humanness would increase with the greater affective transition (Fig.~\ref{fig7}). Our further analysis of AT provided more concrete suggestions for the self-driving algorithm to offer a human-like ride experience for the passenger, e.g., improving passengers' positive affect during the starting stage (Table~\ref{tab3}) and ensuring smoother starting and braking (Fig.~\ref{fig8}). 

Mentalising is a holistic process of inferring about a target agent's beliefs, motivations (i.e., cognitive mentalising), emotions and feelings (i.e., affective mentalising \cite{shamay2010role}), which not only plays a pivotal role in human social interaction \cite{Wu_2020,molapour2021seven} but also is central to human-machine communication \cite{Banks_2021,zhang2022dynamical}. We conjecture that the reason behind the phenomena we just described above (e.g., the relatively lower humanness rating and AT in the AI driver condition) is that the current self-driving algorithm may lack a certain level of mentalising ability (especially affective mentalising ability). For instance, without understanding the emotions and feelings of the passenger and how specific driving behaviour affect the passenger's emotions and feelings particularly (for a similar example of pedestrian-AC interaction, see \cite{liu2021importance}), the self-driving algorithm may not be able to provide passengers with a comfortable and pleasant ride experience as the human driver. More generally, as suggested by the literature \cite{cuzzolin2020knowing}, current AI is yet to fully embrace `hot' cognition (refers to emotional and social cognition; in contrast to `cold' cognition processes, i.e., non-emotional information processing \cite{langley2022theory}), and it is crucial that AI applications should include a mentalising system to help improve human-machine interaction. Ergo, we think it is very likely that imbuing future ACs with artificial mentalising ability will increase their human likeness and thus encourage automated driving to be integrated into human society. And the interdisciplinary collaboration incorporating psychology, neuroscience and computer science is the path we must take to develop such kind of artificial social intelligence with mentalising ability \cite{fu2020can,cao2021can,williams2022supporting}.

\subsection{Limitations and future work}

One could argue that passengers' humanness rating behaviour might not emerge completely after but during the stage. In other words, passengers might make the humanness rating first (which later results in their affective transition) during the road stage before they report post-stage emotions. In response to this questioning of logical rationality, let us go back to the buttress of our computational modelling, i.e., Lewin's field theory. One principle of Lewin's field theory is contemporaneity, which means that the behaviour in a psychological field depends only upon the psychological field `at that time' \cite{lewin1943defining}, i.e., $B_t = f(P_t,E_t)$ \footnote{For the sake of brevity, we ignored the time subscript $t$ in the previous description of Lewin's equation.}. Empirically, a `field at a given time' does not refer to a moment without time extension, but to a certain time period \cite{lewin1943defining} (quite similar to describing the velocity of a point with treating a moment as a certain time period in physics \cite{barker1941frustration}). In our case, it is worth noting that the psychological past and psychological future within a road stage are simultaneous parts of the passenger's psychological field existing at a given time $t$. That is to say, to the size of the passenger's humanness rating behaviour, the whole road stage the passenger rode would be considered as the size of the passenger's psychological field (cf. \cite{lewin1943defining}). It cannot be excluded the possibility that passengers' humanness rating behaviour emerged during the road stage before passengers reported post-stage emotions if one is outside of Lewin's field theory, though the contemporaneity principle of field theory has already sufficed to address this concern. Future work should try to investigate this possibility with a more rigorous and sophisticated experimental design.

There are also several limitations that we should address in the present study. First, we conducted the non-verbal variation of the Turing test in a non-social context where no pedestrians were in the test stages. Thus, neither the AI nor the human driver in our experiment will face the so-called social or moral dilemma (e.g. trolley problem) \cite{bonnefon2016social,shariff2017psychological,awad2018moral,yokoi2020trust}. Given the far-reaching importance of AI ethical decision-making to its social acceptance \cite{caro2022society}, further research on this topic is necessary. Second, due to the capacity of people the event could hold, the number of participants in the current study was limited (68, 68 and 65 effective observations in the first stage, second stage and third stage, respectively). Third, we ignored the inherent differences of passengers, e.g., individual differences in their driving experiences and social cognition (large individual differences have been found in human mentalising ability and social behaviour \cite{wu_fung_mobbs_2019,zhang2021variations,wang2022individual,li2022every}), all of which might affect the generalisation of our results. Hence, validation tests (with a larger sample size, conducted in real-life road situations without safety prompts and under different driving and environmental conditions) would be crucial in future work to test whether our findings will remain. Finally, we only used self-reported scores to measure the emotional experiences of passengers, which limits our adventure towards the neural underpinnings supporting passengers' ascription of humanness in our non-verbal variation of the Turing test. Future studies might uncover this by using physiological measurement (e.g., heart rate, eye-movement entropy, galvanic skin response \cite{dillen2020keep}), mobile electroencephalography (EEG) \cite{Aspinall_2013} (or even combined with mouse-tracking \cite{chen2022resource}) or portable functional near-infrared spectroscopy \cite{Piper_2014,Si_2015}.

\section*{Data and Code Availability Statement}

The data and code used in this paper are available at \url{http://github.com/Das-Boot/bot\_or\_not}.

\section*{Acknowledgements}

We thank Chancheng Zhou for assisting with the experiment. This work was supported by the National Natural Science Foundation of China (32171082), the National Social Science Foundation of China (17ZDA323), the Neuroeconomics Laboratory of Guangzhou Huashang College (2021WSYS002) and Science and Technology Development Fund (FDCT) of Macau [0127/2020/A3].

\beginsupplement

\appendix

\begin{table*}[!ht]
\renewcommand{\arraystretch}{1.3}
\caption{Regression and correlation results}
\label{tabA1}
\begin{subtable}[!h]{0.495\textwidth}
\centering
\subcaption{Direct correlations with $B$.} 
\label{tabA11}
\begin{tabular}{c||ccc}
\hline\hline
Self-reported scores            & \textit{First stage}         & \textit{Second stage}         & \textit{Third stage}         \\ \hline\hline
\multicolumn{1}{l||}{\textit{Pre-study baseline}}  & \multicolumn{1}{l}{} & \multicolumn{1}{l}{} & \multicolumn{1}{l}{} \\
Enjoyment                                & -0.02 (0.86)          & -0.10 (0.41)            & 0.01 (0.95)         \\
Interest                                 & 0.02 (0.90)           & -0.10 (0.40)            & 0.09 (0.46)         \\
Surprise                                 & -0.07 (0.58)          & -0.18 (0.13)            & -0.07 (0.58)          \\
Fear                                     & 0.12 (0.31)           & -0.02 (0.84)            & -0.14 (0.27)          \\
Tension                                  & 0.06 (0.60)           & -0.07 (0.60)            & \textbf{-0.24 (0.05*)} \\
Satisfaction                             & -0.04 (0.75)          & -0.12 (0.34)            & 0.06 (0.65)         \\ \hline\hline
\multicolumn{1}{l||}{\textit{Post-stage}}& \multicolumn{1}{l}{} & \multicolumn{1}{l}{}   & \multicolumn{1}{l}{} \\
Safety                                   & -0.01 (0.92)          & -0.15 (0.21)            & 0.00 (1.00)          \\
Comfort                                  & -0.06 (0.62)          & -0.16 (0.18)            & -0.10 (0.45)          \\
Enjoyment                                & 0.00 (0.99)           & -0.08 (0.52)            & -0.12 (0.34)          \\
Interest                                 & -0.06 (0.64)          & \textbf{-0.28 (0.02*)}  & 0.03 (0.84)         \\
Surprise                                 & -0.01 (0.94)          & \textbf{-0.34 (0.00**)} & -0.03 (0.81)          \\
Fear                                     & 0.09 (0.46)           & \textbf{-0.25 (0.04*)}  & \textbf{-0.26 (0.04*)} \\
Tension                                  & -0.08 (0.52)          & \textbf{-0.27 (0.03*)}  & -0.24 (0.06) \\
Satisfaction                             & 0.03 (0.83)           & \textbf{-0.33 (0.01**)} & -0.08 (0.52)          \\ \hline\hline
\end{tabular}
\end{subtable}
%\\[6pt]
\hspace{\fill}
\begin{subtable}[!h]{0.495\textwidth}
\centering
\subcaption{Regression results with $B$ as dependent variable.}
\label{tabA12}
\begin{tabular}{c||ccc}
\hline\hline
Self-reported scores                     & \textit{First stage}   & \textit{Second stage}    & \textit{Third stage}  \\ \hline\hline
\multicolumn{1}{l||}{\textit{Pre-study baseline}}  & \multicolumn{1}{l}{}  & \multicolumn{1}{l}{}  & \multicolumn{1}{l}{}  \\
Enjoyment                                & -0.05 (0.94)           & -0.28 (0.60)             & \textbf{-1.42 (0.02*)}  \\
Interest                                 & \textbf{1.23 (0.03*)}  & 0.24 (0.68)              & \textbf{1.51 (0.01*)} \\
Surprise                                 & -0.43 (0.25)           & -0.20 (0.64)             & 0.02 (0.98)          \\
Fear                                     & 1.05 (0.13)            & 0.93 (0.26)              & \textbf{2.01 (0.01**)} \\
Tension                                  & -0.53 (0.41)           & -0.12 (0.88)             & \textbf{-2.11 (0.01**)}  \\
Satisfaction                             & -0.58 (0.18)           & 0.42 (0.46)              & 0.36 (0.47)          \\ \hline\hline
\multicolumn{1}{l||}{\textit{Post-stage}}& \multicolumn{1}{l}{}   & \multicolumn{1}{l}{}     & \multicolumn{1}{l}{}  \\
Safety                                   & 0.47 (0.10)            & -0.01 (0.98)             & \textbf{1.09 (0.00***)}          \\
Comfort                                  & -0.37 (0.33)           & 0.20 (0.59)              & -0.25 (0.52)           \\
Enjoyment                                & 0.15 (0.82)            & \textbf{2.05 (0.00**)}   & -0.58 (0.28)           \\
Interest                                 & -1.08 (0.12)           & -1.32 (0.07)             & 0.60 (0.31)          \\
Surprise                                 & -0.17 (0.69)           & -0.63 (0.21)             & 0.31 (0.49)          \\
Fear                                     & 1.67 (0.06)            & -1.38 (0.17)             & -1.73 (0.23)           \\
Tension                                  & -1.71 (0.05)           & -0.12 (0.90)             & 0.22 (0.88)          \\
Satisfaction                             & 0.83 (0.10)            & \textbf{-1.16 (0.03*)}   & -0.43 (0.32)           \\ \hline\hline
\end{tabular}
\end{subtable}
\\[6pt]
\small `$B$' for passengers' humanness rating behaviour. All correlation coefficients in Table~\ref{tabA11} represent Spearman's rank correlation score, and the regression coefficients in Table~\ref{tabA12} were derived from ordinal logistic regressions. The significant effects (* $p < .05$, ** $p < .01$, *** $p < .001$) are bold. All the $p$-values (uncorrected) in parentheses were derived from two-tailed tests.
\end{table*}

\begin{figure*}[!h]
\center
\includegraphics[width=\textwidth]{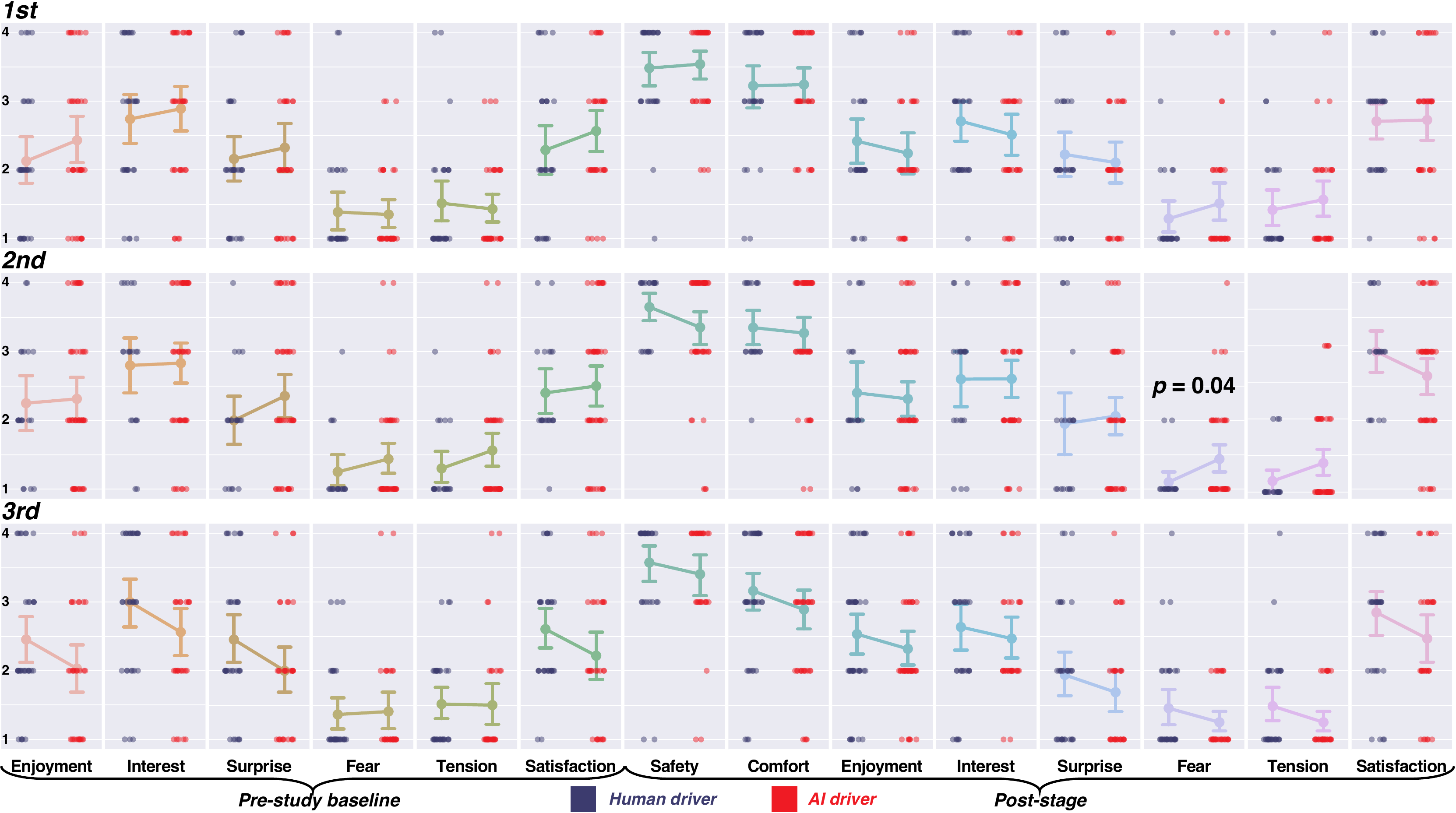}
\caption{The raw self-reported scores (including pre-study baseline emotions, post-stage emotions, safety and comfort) and their mean values, as well as 95\% confidence intervals, differentiated by stimuli (human and AI drivers). Two-tailed Mann-Whitney U tests confirmed that the raw scores under the human driver condition were not significantly different from those under the AI driver condition, except for the fear scores in the second stage (uncorrected $p$ = $0.04$). Notice that pre-study baseline emotions and post-stage emotions were collected using the modified DES-IV on Likert scales from 1-4. Self-reported scores of safety and comfort were rated on an integer scale from 1 to 4, 1 meaning `Not safe (comfortable) at all' and 4 meaning `Very safe (comfortable)'. `1st' for the first stage; `2nd' for the second stage; `3rd' for the third stage. For the first and the second stage, $N\text{ = }68$; for the third stage, $N\text{ = }65$.}
\label{figA1}
\end{figure*}

\ifCLASSOPTIONcaptionsoff
  \newpage
\fi

\newpage

\bibliography{IEEEabrv,manuscript}

\newpage

\begin{IEEEbiography}[{\includegraphics[width=1in,height=1.25in,clip,keepaspectratio]{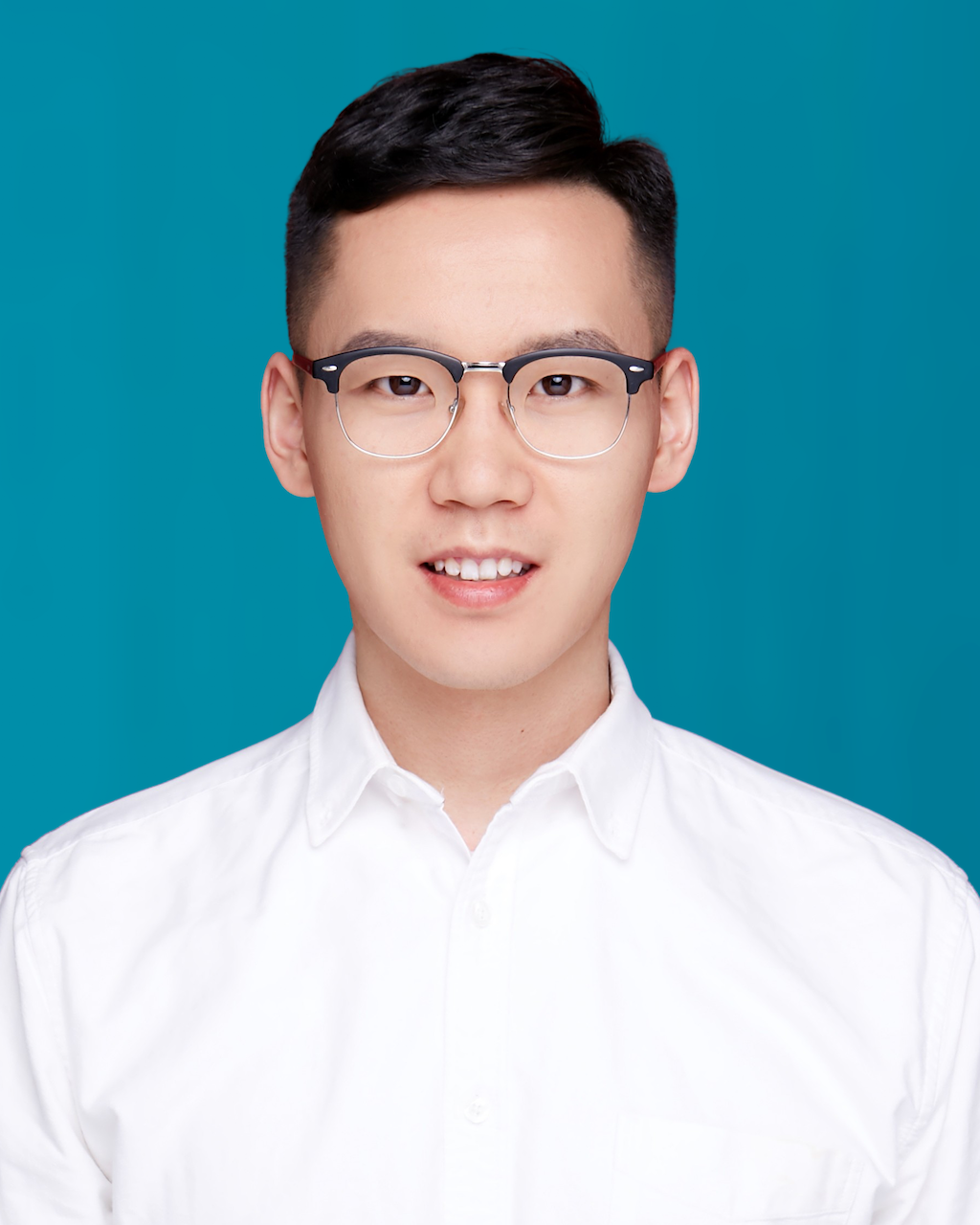}}]{Zhaoning Li}
received a B.E. degree in Information Security in 2016 and an M.E. degree in Software Engineering in 2018, both from Sun Yat-sen University. He is a PhD student at the University of Macau. His research interests include social cognition, social neuroscience, mentalising and artificial social intelligence. Homepage: \href{https://lizhaoning.academia.edu/}{lizhaoning.academia.edu}. Social media: \href{https://twitter.com/lizhn7}{@lizhn7} (Twitter); \href{https://sciences.social/@lizhn7}{@lizhn7@sciences.social} (Mastodon).
\end{IEEEbiography}

\vspace{6pt}

\begin{IEEEbiography}[{\includegraphics[width=1in,height=1.25in,clip,keepaspectratio]{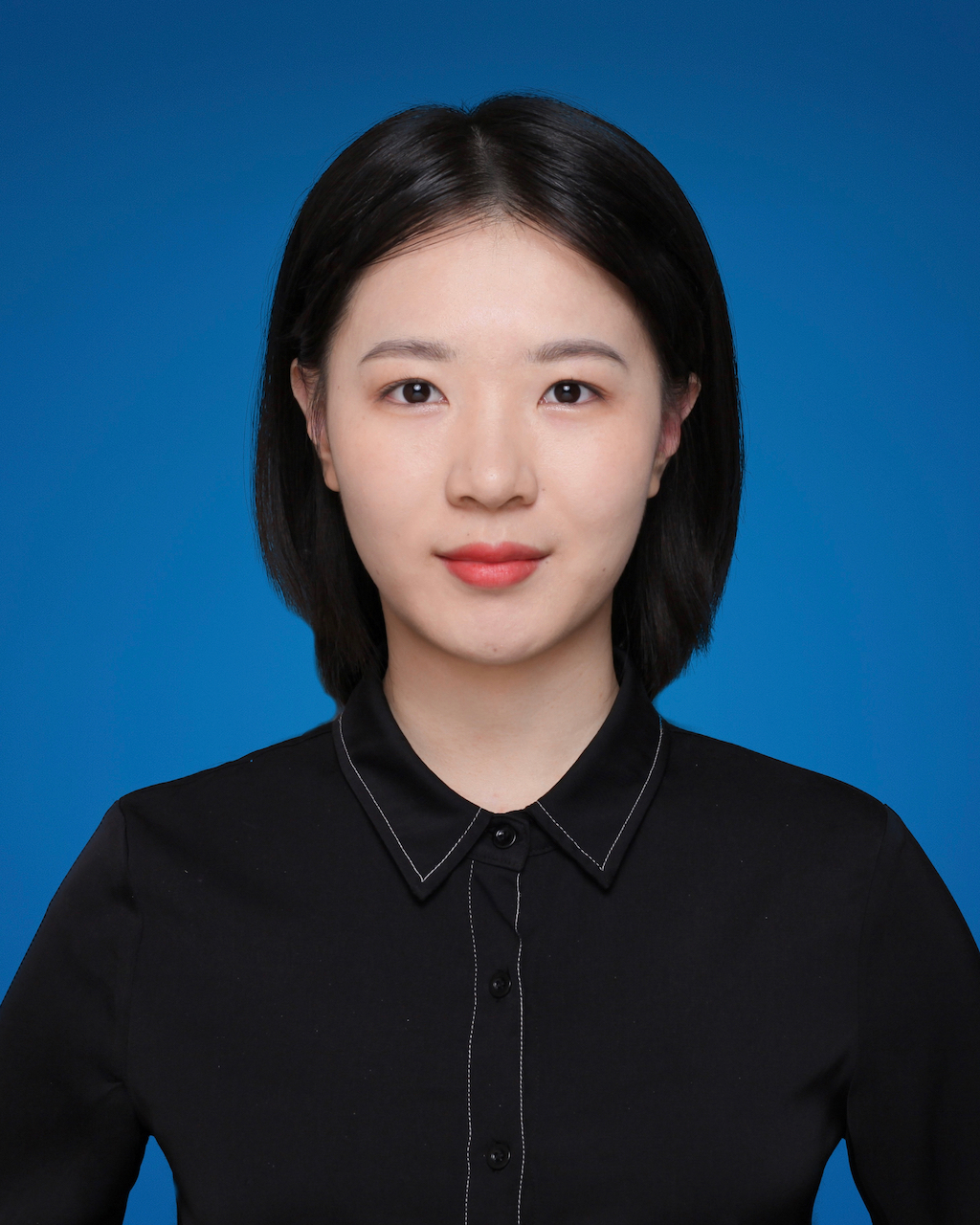}}]{Qiaoli Jiang}
received her master's degree in psychology from Sun Yat-Sen University in China in 2021. She specialises in perception, well-being and emotional processes during social interaction.
\end{IEEEbiography}

\vspace{6pt}

\begin{IEEEbiography}[{\includegraphics[width=1in,height=1.25in,clip,keepaspectratio]{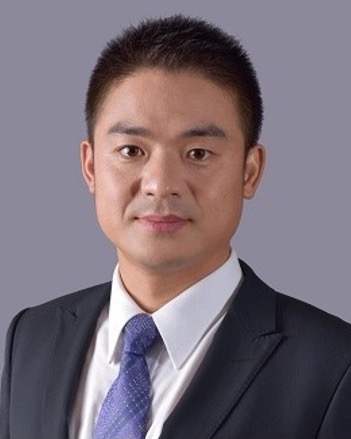}}]{Zhengming Wu}
is the Director of the Guangzhou Intelligent Connected Vehicle Pilot Zone Operations Centre. He is mainly responsible for road test supervision, policy planning consultation, and business cooperation. He has ten years of equity investment experience, with rich experience in corporate operation, capital investment and financing.
\end{IEEEbiography}

\vspace{6pt}

\begin{IEEEbiography}[{\includegraphics[width=1in,height=1.25in,clip,keepaspectratio]{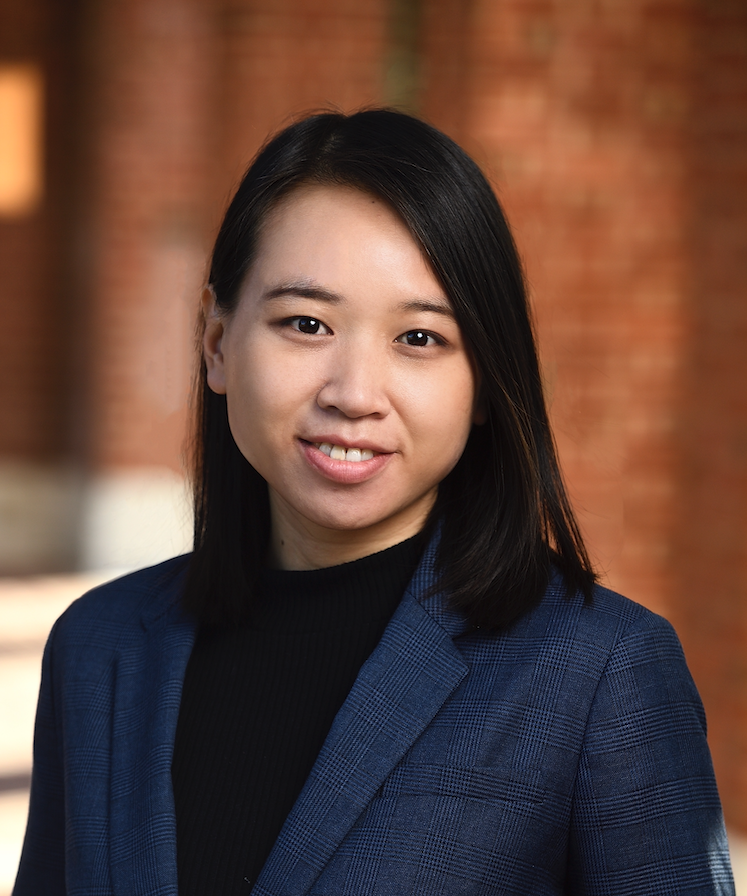}}]{Anqi Liu}
received a PhD degree in computer science from the University of Illinois at Chicago, United States, in 2018. She is an Assistant Professor at the Department of Computer Science, Johns Hopkins University. Her research interest lies in machine learning for trustworthy AI. She is broadly interested in developing principled machine learning algorithms for building more reliable, trustworthy, and human-compatible AI systems in the real world. Homepage: \href{https://anqiliu-ai.github.io/}{anqiliu-ai.github.io}. Twitter: \href{https://twitter.com/anqi_liu33}{@anqi\_liu33}.
\end{IEEEbiography}

\vspace{6pt}

\begin{IEEEbiography}[{\includegraphics[width=1in,height=1.25in,clip,keepaspectratio]{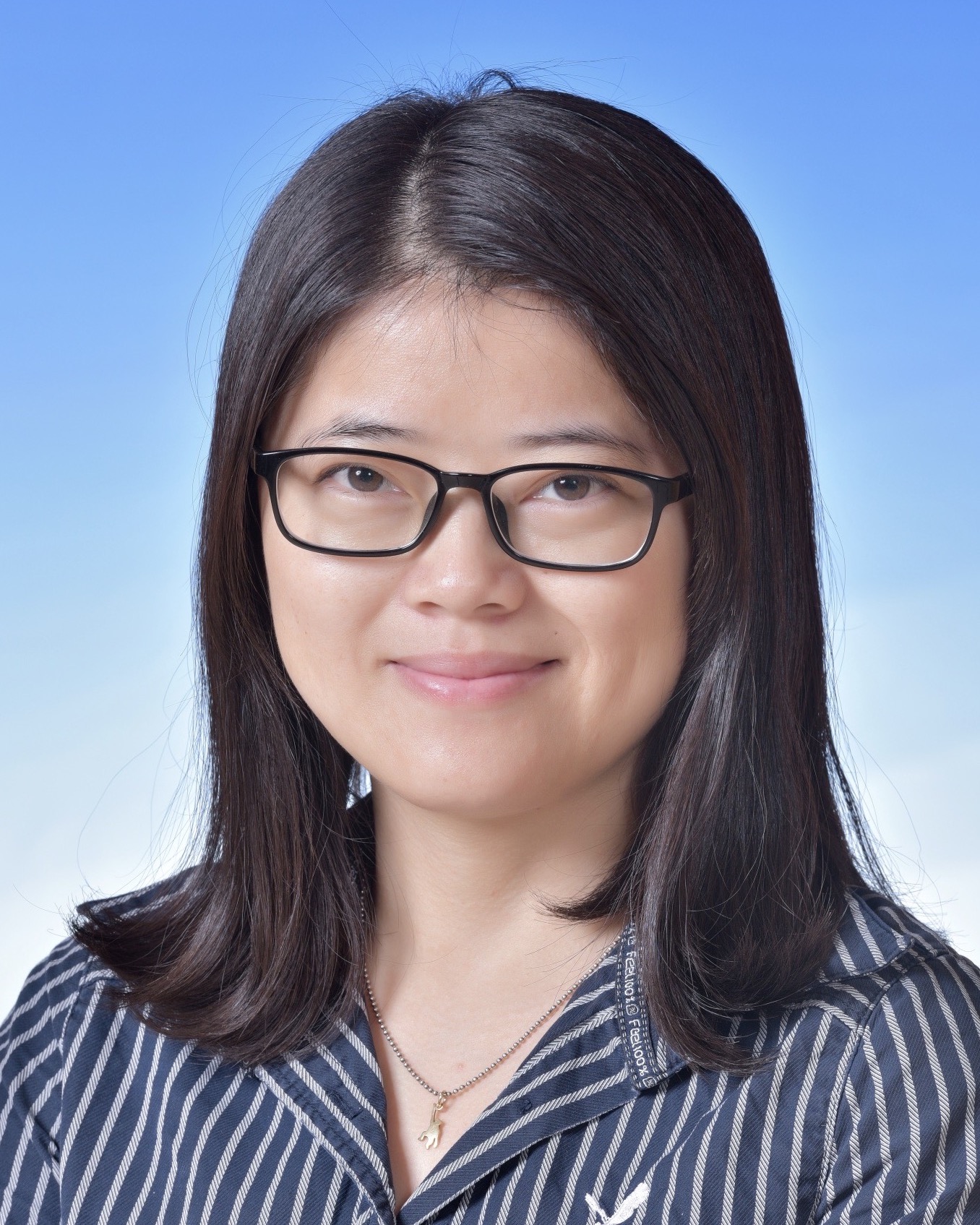}}]{Haiyan Wu}
received a PhD degree in cognitive neuroscience from Beijing Normal University, China, in 2013. She is an Assistant Professor at the Centre for Cognitive and Brain Sciences, University of Macau. She is interested in exploring the mechanisms that describe and/or affect behavioural and neural responses during emotion and decision making (both for human and human-AI interaction) in social interaction. Her interest also lies in the area of the interactions between social relations cognition and languages. Homepage: \href{https://andlab-um.com/}{andlab-um.com} (lab); \href{https://haiyanwu.wixsite.com/haiyanwu}{haiyanwu.wixsite.com/haiyanwu} (personal). Twitter: \href{https://twitter.com/ANDlab3}{@ANDlab3} (lab); \href{https://twitter.com/Wu3Haiyan}{@Wu3Haiyan} (personal).
\end{IEEEbiography}

%\vspace{9pt}

\begin{IEEEbiography}[{\includegraphics[width=1in,height=1.25in,clip,keepaspectratio]{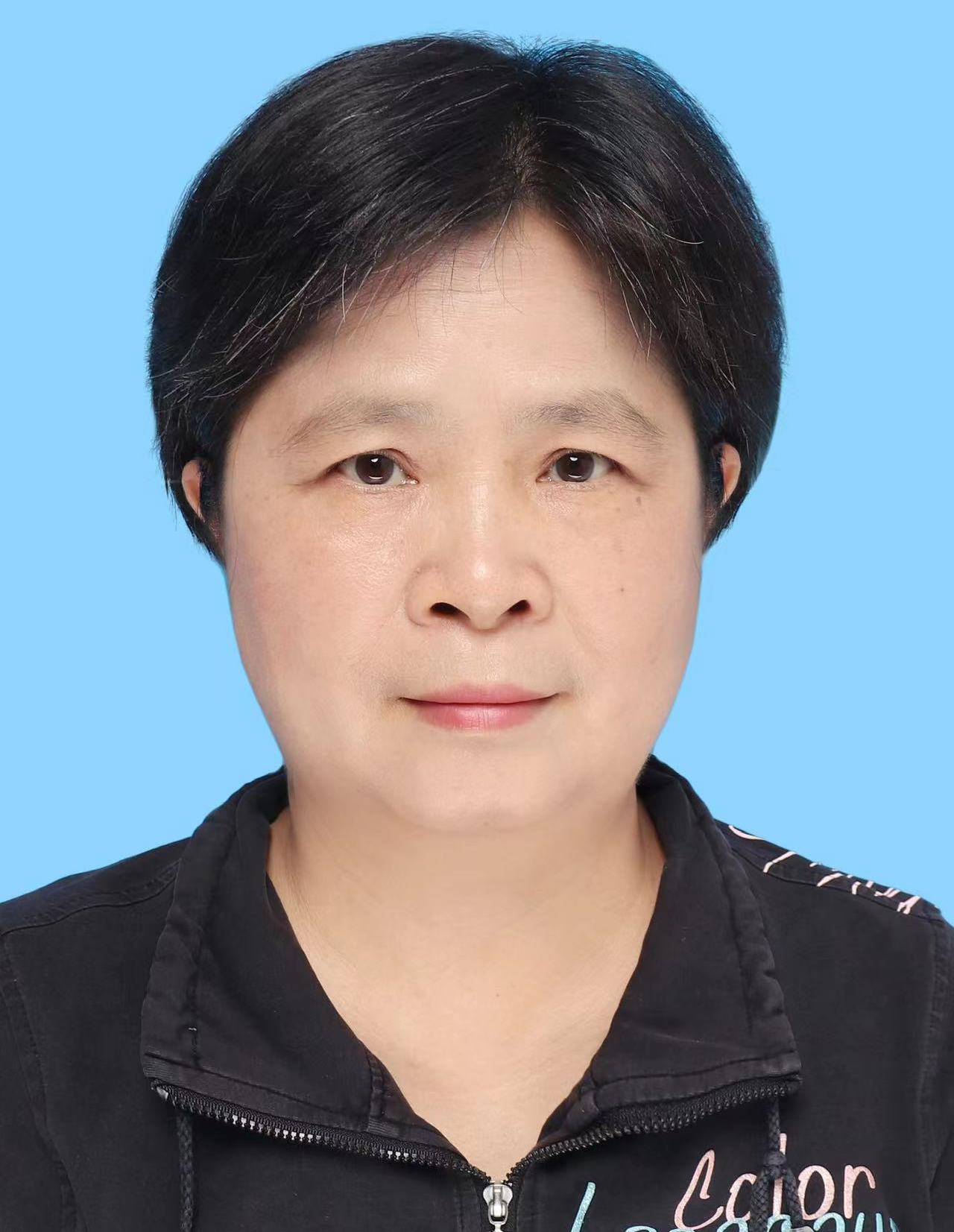}}]{Miner Huang}
received a PhD degree in psychology from Capital Normal University, China, in 2001. She is a Professor at the Department of Psychology, Sun Yat-sen University. Her research interests include interpersonal and social sharing of emotions, emotion regulation and emotional labour.
\end{IEEEbiography}

\vspace{18pt}

\begin{IEEEbiography}[{\includegraphics[width=1in,height=1.25in,clip,keepaspectratio]{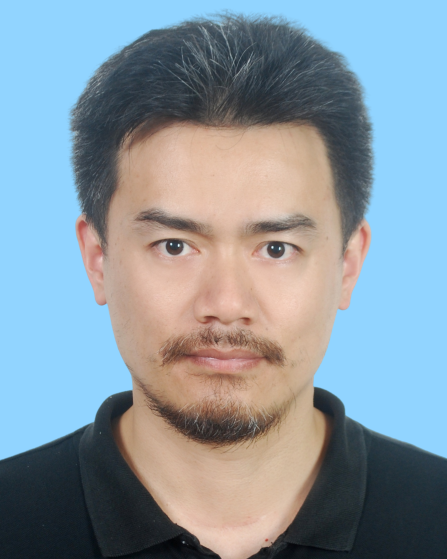}}]{Kai Huang}
received a PhD degree in electronic engineering from ETH Zurich, Switzerland, in 2010. He is a Professor at the School of Computer Science and Engineering, Sun Yat-sen University. His research interests include techniques for the analysis, design, and optimization of embedded/CPS systems, particularly in the automotive, medical, and robotic domains. Homepage: \href{https://www.usilab.cn/hk/}{usilab.cn/hk/} (personal); \href{https://www.usilab.cn/}{usilab.cn} (lab).
\end{IEEEbiography}

\vspace{4.5pt}

\begin{IEEEbiography}[{\includegraphics[width=1in,height=1.25in,clip,keepaspectratio]{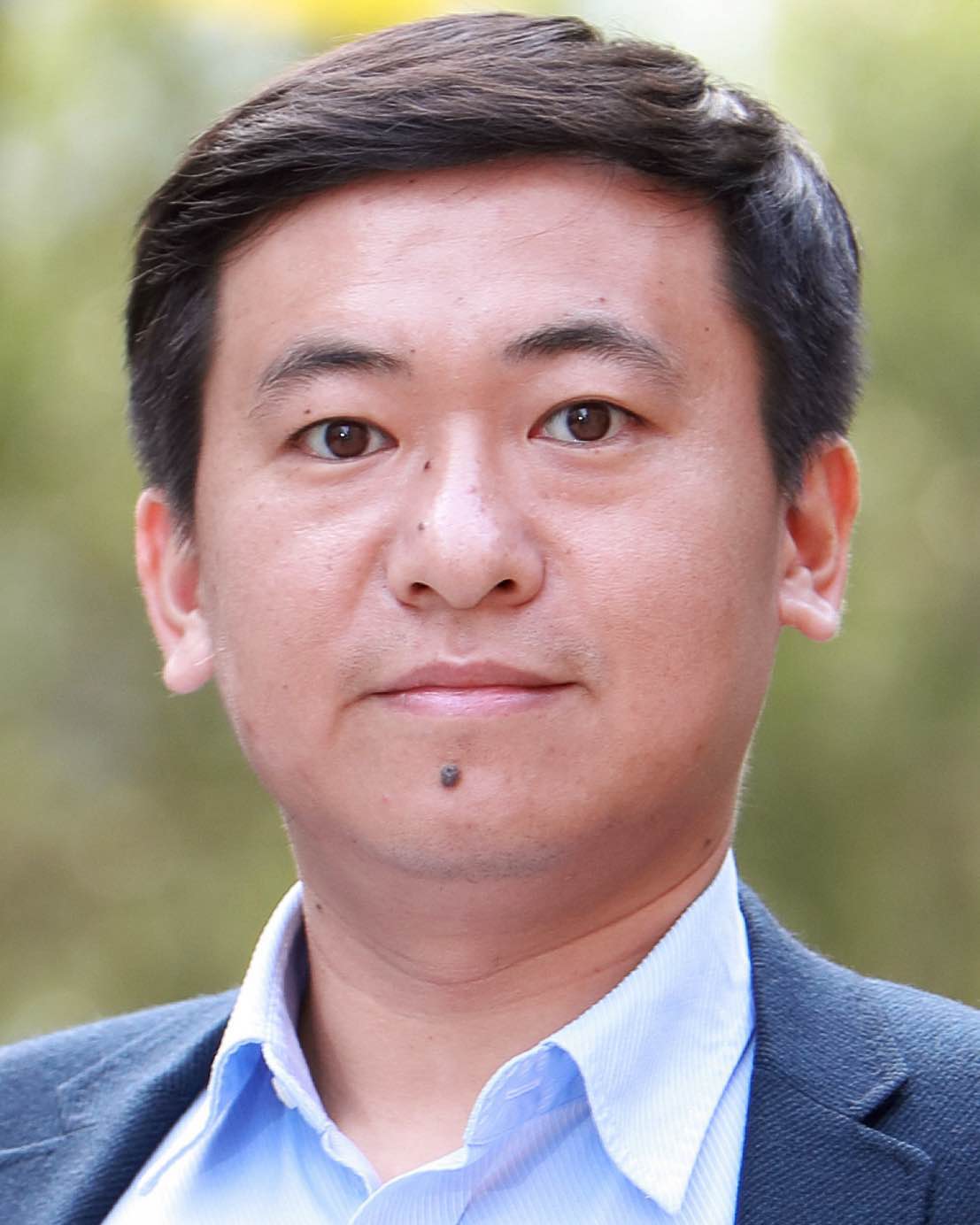}}]{Yixuan Ku}
received a PhD degree in biomedical engineering from Tsinghua University, China, in 2010. He is a Professor at the Department of Psychology, Sun Yat-sen University. His research interests include neural mechanisms of memory and emotion and the interdisciplinary research of neuroscience and artificial intelligence. Homepage: \href{https://psy.sysu.edu.cn/teacher/851}{psy.sysu.edu.cn/teacher/851} (personal); \href{https://sysumelab.com/}{sysumelab.com}. Twitter: \href{https://twitter.com/KuYixuan}{@KuYixuan}.
\end{IEEEbiography}

\vfill

\end{document}